\documentclass[%
reprint,
superscriptaddress,
preprintnumbers,
nofootinbib,
amsmath,amssymb,
aps,
prd,
]{revtex4-2}

\usepackage{graphicx}% Include figure files
\usepackage{dcolumn}% Align table columns on decimal point
\usepackage{bm}% bold math
\usepackage[colorlinks, linkcolor=blue, pdfborder={0 0 0}, breaklinks=true]{hyperref}

\usepackage{xcolor}
\usepackage{soul}

\newcommand{\mpch}{\>h^{-1}{\rm {Mpc}}}
\newcommand{\hmpc}{\>h\mathrm{Mpc}^{-1}}

\newcommand{\rmHI}{{\rm HI}}
\newcommand{\rmfg}{{\rm fg}}

\newcommand{\rmMHz}{{\rm MHz}}
\newcommand{\rmns}{{\rm N}}
\newcommand{\rmtrue}{{\rm true}}
\newcommand{\rmrec}{{\rm rec}}
\newcommand{\mnras}{{MNRAS}}
\newcommand{\jcap}{{JCAP}}
\newcommand{\pasp}{{PASP}}
\newcommand{\apjs}{{ApJS}}
\newcommand{\aap}{{A\&A}}
\newcommand{\aaps}{{A\&AS}}

\newcommand{\apjl}{{ApJL}}

\begin{document}

% \preprint{APS/123-QED}

\title{21-cm foreground removal using AI and frequency-difference technique}% Force line breaks with \\

\author{Feng Shi}
\altaffiliation{fshi@xidian.edu.cn}
\affiliation{School of Aerospace Science And Technology, Xidian University, Xi'an 710126, P.R. China}
\affiliation{Peng Cheng Laboratory, No.2, Xingke 1st Street, Shenzhen 518000, P.R. China}

\author{Haoxiang Chang}
\affiliation{School of Aerospace Science And Technology, Xidian University, Xi'an 710126, P.R. China}

\author{Le Zhang}
\affiliation{School of Physics and Astronomy, Sun Yat-sen University, 2 Daxue Road, Tangjia, Zhuhai, 519082, P.R. China}
\affiliation{CSST Science Center for the Guangdong-Hong Kong-Macau Greater Bay Area, Zhuhai 519082, P.R. China}
\affiliation{Peng Cheng Laboratory, No.2, Xingke 1st Street, Shenzhen 518000, P.R. China}

\author{Huanyuan Shan}
\affiliation{Shanghai Astronomical Observatory (SHAO), Nandan Road 80, Shanghai 200030, China}
\affiliation{University of Chinese Academy of Sciences, Beijing 100049, P.R. China}
\author{Jiajun Zhang}
\affiliation{Shanghai Astronomical Observatory (SHAO), Nandan Road 80, Shanghai 200030, China}
\affiliation{University of Chinese Academy of Sciences, Beijing 100049, P.R. China}
\author{Suiping Zhou}
\affiliation{School of Aerospace Science And Technology, Xidian University, Xi'an 710126, P.R. China}

\author{Ming Jiang}
\affiliation{National Key Laboratory of Radar Signal Processing, Xidian University, Xi'an 710126, P.R. China}
\author{Zitong Wang}
\affiliation{School of Aerospace Science And Technology, Xidian University, Xi'an 710126, P.R. China}

\date{\today}% It is always \today, today,

\begin{abstract}
The deep learning technique has been employed in removing foreground contaminants from 21-cm intensity mapping, but its effectiveness is limited by the large dynamic range of the foreground amplitude. In this study, we develop a novel foreground removal technique grounded in U-Net networks. The essence of this technique lies in introducing an innovative data preprocessing step specifically, utilizing the temperature difference between neighboring frequency bands as input. Combining with the frequency difference, we refer to our method as the UNet-fd (UNet frequency-difference), where the U-Net structure is the same as that in \texttt{deep21}. Based on our tests, we demonstrate that this frequency-difference preprocessing technique can substantially reduce the dynamic range of foreground amplitudes by approximately two orders of magnitude. This reduction proves to be highly advantageous for the U-Net foreground removal. We observe that the HI signal can be reliably recovered, as indicated by the cross-correlation power spectra showing unity agreement at the scale of $k\lesssim0.3\hmpc$ in the absence of instrumental effects. Moreover, accounting for the systematic beam effects, our reconstruction displays consistent auto-correlation and cross-correlation power spectrum ratios at the $1\sigma$ level across scales $k \lesssim 0.1\hmpc$, with only a 10\% reduction observed in the cross-correlation power spectrum at $k\simeq0.2\hmpc$. The effects of redshift-space distortion are also reconstructed successfully, as evidenced by the quadrupole power spectra matching with the target truth. In order to test how thermal noise affects the performance of our method, we simulated various white noise levels in the map. This shows the mean cross-correlation ratio $\bar{R}_\mathrm{cross} \gtrsim 0.8$ when the level of the thermal noise is smaller than or equal to that of the HI signal. In comparison, our method outperforms the traditional Principal Component Analysis (PCA) method. The PCA-derived cross-correlation ratios are underestimated by around 60\%. We conclude that the proposed frequency-difference technique can significantly enhance network performance by reducing the amplitude range of foregrounds and aiding in the prevention of HI loss.

\end{abstract}

\maketitle

\section{Introduction} \label{sec:intro}
In the next few decades, 21-cm cosmology has the potential to dramatically improve our knowledge of the universe, providing a fully three-dimensional mapping of the HI spectral line from the dark ages \citep{1990MNRAS.247..510S,2004PhRvL..92u1301L},  to the cosmic dawn \citep[e.g.][]{2008ApJ...680..962L,2013JCAP...09..014C} and the postreionization \citep[e.g.][]{2016JApA...37...26S, 2016MNRAS.459..151C,2017JCAP...04..001C, 2017MNRAS.471.3112M}. One promising technique for surveying the low-redshift large-scale structure (LSS) of the Universe is 21-cm intensity mapping (IM), which involves measuring the integrated HI emission lines that originate from unresolved sources in large-volume portions of the sky\citep{2001JApA...22...21B,2004MNRAS.355.1339B,2008PhRvL.100i1303C,2009astro2010S.234P,2009MNRAS.397.1926W,2010ApJ...721..164S}. While this is conceptually similar to the traditional galaxy redshift survey, it differs significantly in that 21-cm IM is sensitive to all sources of emission rather than just cataloging the brightest galaxies. This allows for mapping LSS through the redshift desert and beyond (such as $1 \lesssim z \lesssim 6$, where optical spectroscopy is challenging~\citep{2020PASP..132f2001L}.

A large number of interferometric and imaging radio telescopes are currently conducting and will continue to carry out 21-cm IM observations, such as CHIME (Canadian Hydrogen Intensity Mapping Experiment) \citep{2023ApJ...947...16A,2022ApJS..261...29C}, Tianlai~\citep{2011SSPMA..41.1358C, 2015ApJ...798...40X, 2021MNRAS.506.3455W,2022MNRAS.517.4637P}, BINGO (Baryon acoustic oscillations In Neutral Gas Observations)~\citep{2013MNRAS.434.1239B,2022BINGO}, MeerKAT (The South African Square Kilometer Array Pathfinder)~\citep{2016mks..confE..32S, 2021MNRAS.505.3698W}, HIRAX (The Hydrogen Intensity and Real-time Analysis eXperiment)~\citep{2016SPIE.9906E..5XN}, and the SKA (The Square Kilometer Array)~\citep{2015aska.confE..19S}. So far, most studies have focused on 21-cm IM in cross-correlation with galaxy surveys~\citep{2010Natur.466..463C, 2013ApJ...763L..20M, 2013MNRAS.434L..46S,2018MNRAS.476.3382A, 2022MNRAS.510.3495W,2023MNRAS.518.6262C}. More recently, there has been a successful detection of the HI autocorrelation power spectrum on small scales using MeerKAT~\citep{2023arXiv230111943P}.

However, the biggest challenge for upcoming observational data analysis is the presence of foreground contaminants, such as the Galactic synchrotron and free-free emissions and extragalactic radio point sources, which are about 4 orders of magnitude brighter than the HI signal~\citep{2002ApJ...564..576D,2003MNRAS.346..871O,2005ApJ...625..575S,2008MNRAS.388..247D}. 

Over the years, a variety of signal separation algorithms have been consistently proposed. These algorithms can be categorized as either 'blind' or 'non-blind', depending on whether prior knowledge of the signal, foreground, or noise is required for the separation process. A comprehensive overview of current techniques for mitigating 21-cm foreground is provided by~\cite{Liu:2019awk}. For instance, Principal Component Analysis (PCA)~\citep{2011PhRvD..83j3006L}, Singular Value Decomposition (SVD)~\citep{Switzer:2015ria, Zuo:2022wra}, Independent Component Analysis (ICA)\citep{2012MNRAS.423.2518C, Cunnington21}, and Generalized Morphological Component Analysis (GMCA)~\citep{2013MNRAS.429..165C} represent blind (or 'semi-blind') methods commonly employed in 21-cm intensity mapping. Conversely, non-blind methods such as Generalized Needlet Internal Linear Combination (GNILC)~\citep{2016MNRAS.456.2749O, Marins22}, Gaussian Process Regression (GPR)~\citep{2012MNRAS.423.2518C, Kern:2020kky}, and the Karhunen-Loeve Transform~\citep{Shaw:2013wza}, are also extensively cited in the literature. Moreover, an approach known as `foreground avoidance' is utilized, where Fourier modes measured above the so-called `wedge' region in $(k_\perp, k_\parallel)$-space are dominantly associated with the 21-cm signal~\citep{2012ApJ...752..137M}. In addition, the Bayesian framework \citep{Ghosh:2015fxa, 2016ApJS..222....3Z, Sims:2019iyz, 2023MNRAS.520.4443B} allows for the simultaneous inference of 21-cm signals and foreground signals under certain a {\it priori} assumptions.

More recently, there has been a growing interest in exploring deep learning techniques for foreground removal, such as by using deep neural networks with U-Net architecture~\citep{2021JCAP...04..081M, 2022ApJ...934...83N}. It has been demonstrated that the U-Net is effective in recovering the HI fluctuation signal, provided that certain preprocessing steps, such as PCA, are undertaken. This initial preprocessing aims to firstly eliminate the dominated foreground components. Subsequently, the U-Net is applied to further eliminate any remaining foreground residuals. However, it should be noted that employing PCA-based preprocessing, which involves subtracting a specific number of PCA modes, may result in an undesired loss of the HI signal.

All of these methods essentially assume that the statistical characteristics of the foreground, as well as the smoothness characteristics of the foreground emissions across the frequencies, differ from those of the 21-cm signal. As a result, they can effectively remove most of the foreground components\citep{2006ApJ...650..529W,2008MNRAS.391..383G,2008MNRAS.389.1319J}. However, because the HI signal vector and individual foreground component vectors are not perfectly orthogonal, the process of subtracting the foreground leads to a simultaneous subtraction of the signal itself. Meanwhile, achieving a clean reconstruction of the 21cm signal is further complicated by the presence of contamination from non-smoothed foreground components~\citep{2015ApJ...815...51S,2020MNRAS.499.4613S, 2020MNRAS.495.1788A, 2023MNRAS.523.2453C}.

In order to avoid signal loss and to more cleanly eliminate non-smoothed foreground components, we propose a new method -- a technique that utilizes the temperature difference of the sky maps in neighboring frequency bands. Since the HI and the foreground have different frequency correlations, this is expected to significantly reduce the dynamic range of the foreground while avoiding HI signal loss. We will demonstrate that the U-Net performs more reliably in recovering large-scale HI information when trained using the frequency-difference map as input.

% maintain the neighboring-band HI signals while suppressing the dynamical range of  foreground temperature. A neural network like U-Net still needs to be used to extract the required signal. The neighboring-bands difference is notably helpful for keeping the HI signal rather than using PCA as the preprocessing. 

This paper is organized as follows. In Sec.~\ref{sec:data} we introduce our simulation. In Sec.\ref{sec:method}, we present our foreground removal method, including the U-Net structure and the frequency-difference preprocessing technique.  We verify our method and present the results in Sec.\ref{sec:result}, and summarize our main findings in Sec.\ref{sec:summary}.

\section{datasets}
\label{sec:data}

This section describes our simulated 21-cm intensity maps, which are generated using {\tt CRIME}\footnote{http://intensitymapping.physics.ox.ac.uk/CRIME.html}~\citep{2014MNRAS.444.3183A} in {\tt HEALPIX} pixelization scheme~\citep{2005ApJ...622..759G}. The code is designed to generate the main components of the 21-cm intensity mapping observations, including the cosmological 21-cm signal, foreground emissions, and instrumental noise. Here, we briefly introduce the simulation method and refer the reader to the references for more details.

\begin{figure*}
    \centering
    \includegraphics[trim=0cm 0cm 0cm 0cm, clip=True,width=1.8\columnwidth]{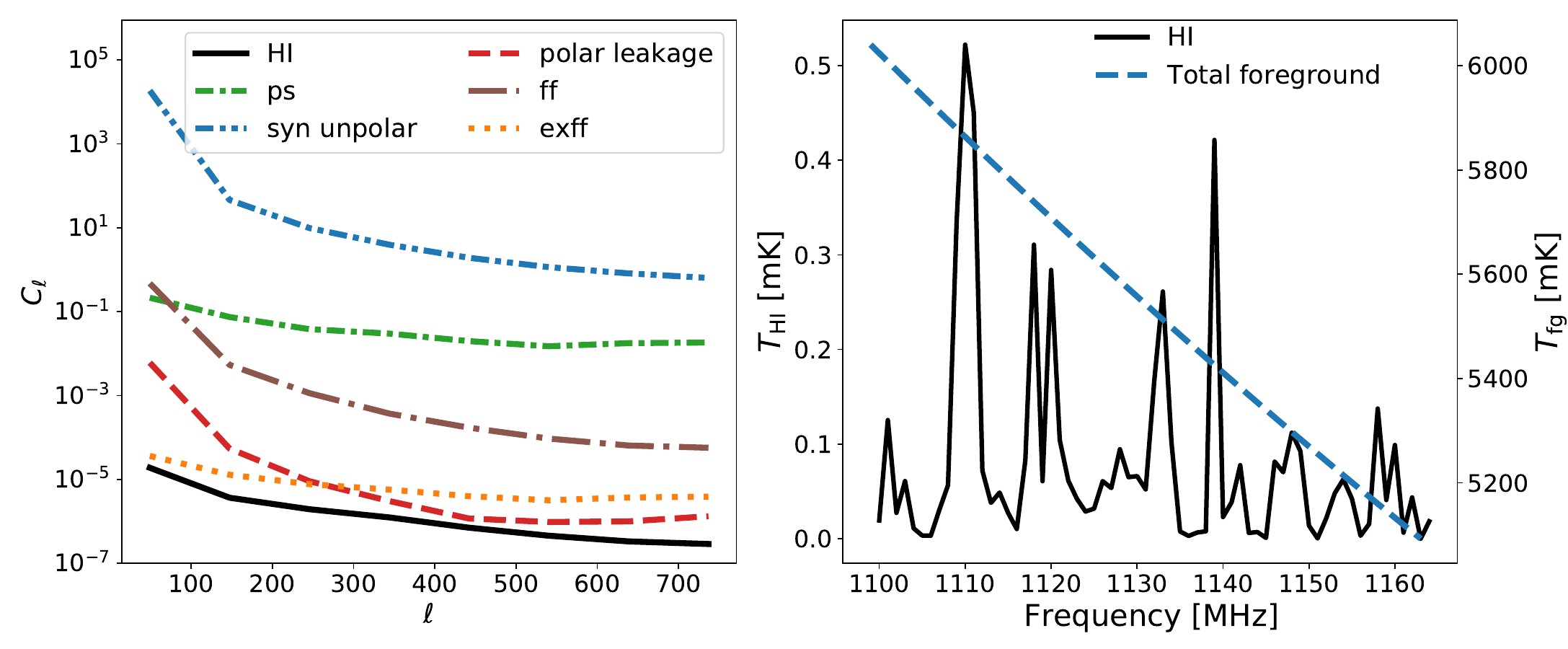}
    \caption{{\it Left}: a comparison of the angular power spectrum among different simulated components at a frequency of $\nu=1110$ MHz. Results are computed for a randomly selected patch from the full sky, which has been divided into 192 patches. The cosmological HI is represented by the black solid line, while other components of the foreground are depicted by colored dashed or dotted lines, as indicated. {\it Right}: the brightness temperature as a function of frequency for a randomly selected sky pixel. The black solid line corresponds to the cosmological HI, while the blue dashed line represents the total foreground. For a clear comparison, note that the amplitude scales are different for the foreground and HI components. }
    \label{fig:ClT_data}
\end{figure*}

\begin{figure*}
    \centering
    \includegraphics[trim=0cm 0cm 0cm 0cm, clip=True,width=2.1\columnwidth]{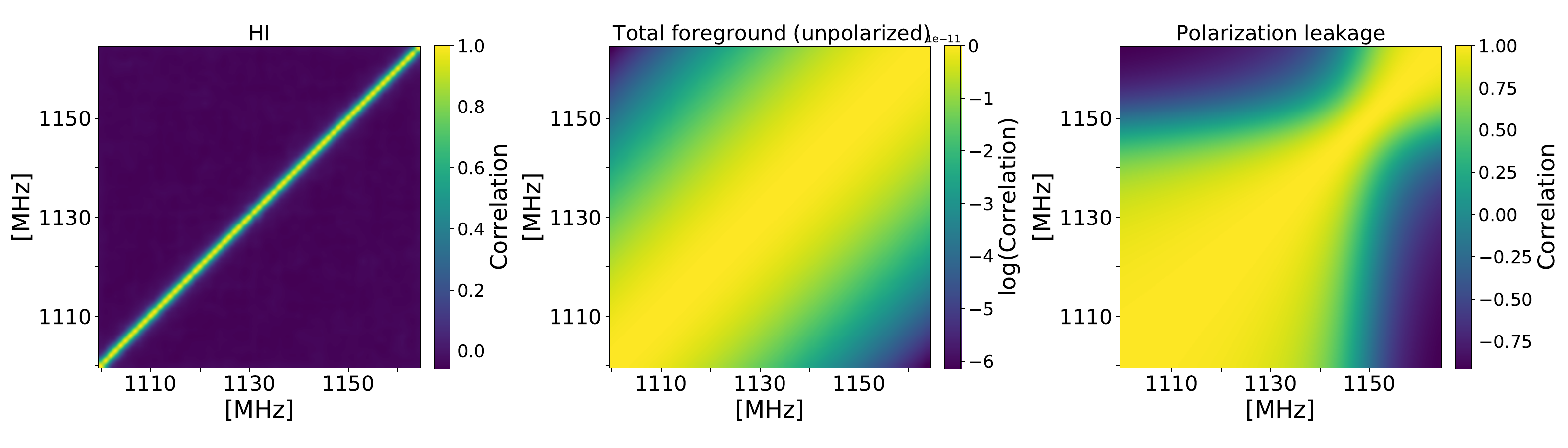}
    \caption{Frequency-frequency correlation matrices for cosmological HI (left), total unpolarized foreground (center), and polarization leakage (right). Please note that the correlation coefficients of the unpolarized foreground are displayed in logarithmic values.}
    \label{fig:CorrMatrix}
\end{figure*}

\begin{figure}
    \centering
    \includegraphics[trim=0cm 0cm 0cm 0cm, clip=True,width=1.\columnwidth]{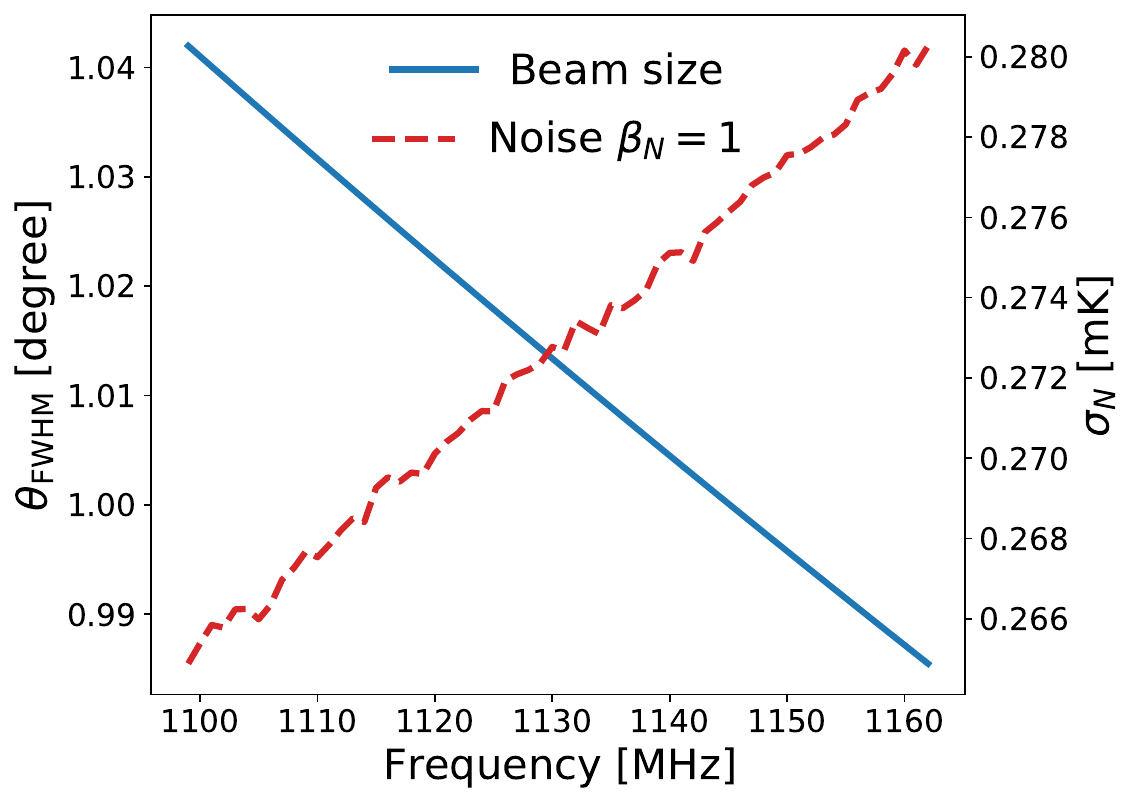}
    \caption{Beam size (blue solid) and thermal noise (red dashed) as a function of frequency, which includes 64 channels. The noise is plotted for the case with $\beta_N=1$ according to Eq.~(\ref{eq:beta_ns}.})
    \label{fig:bN_fre}
\end{figure}

\subsection{Cosmological HI signal}
\label{sec:cosmsignal}

The observed effective brightness temperature of the HI signal, $T_{\rm HI}$, is related to the HI overdensity $\delta_\mathrm{HI}$ at a given sky position $\hat{\bm{n}}$ and redshift $z$ as follows  \citep[e.g.][]{2015ApJ...803...21B}:
\begin{equation}\label{eq:Tb1}
    T_{\rm HI}(\hat{\bm{n}},z) = \bar{T}_\mathrm{HI}(z) (1+\delta_\mathrm{HI}(\hat{\bm{n}},z))\,.
\end{equation}
Here the mean brightness temperature can be expressed as 
\begin{equation}\label{eq:Tb2}
    \centering
    \bar{T}_\mathrm{HI}(z) = 190.55\frac{\Omega _b h(1+z)^2x_{\rm HI}(z)}{\sqrt{\Omega _m(1+z)^3+\Omega _\Lambda }}~{\rm mK}\,,
\end{equation}
where $h$ represents the reduced Hubble parameter, defined as $h=H_0 /(100 \mathrm{~km} \mathrm{~s}^{-1} \mathrm{Mpc}^{-1})$, $x_\mathrm{HI}$ stands for the neutral hydrogen mass fraction relative to the total baryons, and $\Omega_b$, $\Omega_m$, and $\Omega_\Lambda$ correspond to the present fractions of baryon, total matter, and dark energy densities, respectively.

In {\tt CRIME}, the HI overdensity $\delta_\mathrm{HI}$ and velocity perturbations are first generated by employing a log-normal model on a Gaussian realization of the dark matter density field within a Cartesian box. The results are then projected onto spherical sky shells and pixelated to create maps of the 21-cm brightness temperature across various frequency bands, each corresponding to a specific redshift from the observer. The relationship between the brightness temperature and the underlying HI overdensity is established through Eqs.~(\ref{eq:Tb1}) and ~(\ref{eq:Tb2}), assuming $x_\mathrm{HI}(z) = 0.008(1+z)$.

Meanwhile, the redshift-space distortion (RSD) is implemented by perturbing the cosmological redshift of each cell with the redshift distortion $\Delta z_\mathrm{RSD} = (1+z)v_r/c$, where $v_r$ is the line-of-sight velocity of each cell, and $c$ denotes the speed of light.

In this study, we utilize a map with $N_{\rm side}=256$, corresponding to an angular resolution of 13.73 arcmin. We consider the frequency range spanning from 1100 to 1164 MHz, which corresponds to a redshift range between 0.29 and 0.22. Subsequently, the simulated full sky is divided into 192 sky patches, each having dimensions of $64^3$. These patches cover approximately $214.86~ \mathrm{deg}^2$ of the sky and have a volume equivalent to that of a box with a length of about $187 \mpch$.

The cosmological parameters are set as follows: $\Omega_m=0.3$, $\Omega_\Lambda=0.7$, $\Omega_b=0.049$, $h=0.67$, and $\sigma_8=0.8$.

\subsection{Foregrounds}\label{sec:fgrd}
In the simulations, we consider four main foreground components: Galactic synchrotron emission, extragalactic point sources, Galactic and extragalactic free-free emissions. Given the distinct distribution characteristics of anisotropy and isotropy, the simulation for generating the foreground combines two different methods. The Galactic synchrotron emission originates from the accelerated motion of energetic charged particles dispersed in the Galactic magnetic field. It exhibits a highly anisotropic angular structure, with a steep increase in brightness temperature toward the Galactic plane. To produce such a structure, the Galactic synchrotron emission is then simulated by extrapolating the brightness temperature $T_{\rm Haslam}(\hat{\bm{n}})$ from the Haslam map \citep{1982A&AS...47....1H} to the relevant frequencies. This is achieved using a direction-dependent spectral index $\beta(\hat{\bm{n}})$ \citep{2013A&A...553A..96D}, expressed as
\begin{equation}
\centering
    T_{\rm syn,0}(\nu,\hat{\bm{n}})=T_{\rm Haslam}(\hat{\bm{n}}) \left(\frac{\rm 408~MHz}{\nu}\right) ^{\beta(\hat{\bm{n}})}\,.
\label{eq:synchrotron0}
\end{equation}
Because of the low resolution of the Haslam map (approximately $0.85^\circ$), the subscript 0 here denotes the interpolation results at larger scales. As Galactic synchrotron anisotropy on small scales cannot be obtained from the low-resolution Haslam sky map, we simulate the small-scale synchrotron component by generating the isotropic structure of synchrotron emission through Gaussian realizations of the angular power spectra, as modeled by~\citep{2005ApJ...625..575S}, 

% Regarding the point sources and free-free emissions, as well as the small-scale synchrotron component, the foregrounds are generated through Gaussian realizations of the angular power spectra as modeled by~\citep{2005ApJ...625..575S},  
%
\begin{equation}
\centering
    {C}_\ell(v_1,v_2)=A \left(\frac{\ell_{\rm ref}}{\ell}\right)^\beta
                     \left(\frac{v_{\rm ref}^2}{v_1v_2}\right)^\alpha
                     {\rm exp}\left(-\frac{{\rm log}^2{(v_1/v_2)}}{2\xi^2}\right)\,,
	\label{eq:powerspectrum}
\end{equation}
where $A$ represents the amplitude of the power spectrum, and $\xi$ denotes the frequency-space correlation length for a given foreground component. We employ the reference scale $\ell_{\rm ref} = 1000$ and the reference frequency $\nu_{\rm ref} = 130$ MHz. In order to confine the synchrotron simulation to small scales, the random realizations are rescaled to match the resolution scale overlapping with that of the Haslam map before they are incorporated into the final Galactic synchrotron map. Additionally, Eq.~(\ref{eq:powerspectrum}) is also used to simulate free–free emission and point sources. The model parameter values are drawn from~\cite{2005ApJ...625..575S} and are provided in Table~\ref{tab:santos}. Note that, the Haslam map used in the CRIME contains residual striping and point-source artifacts, which could lead to inadvertent duplication of point sources in the simulations. 
This concern can be alleviated by the utilization of an improved version of the Haslam map, which was reprocessed by \citep{2015MNRAS.451.4311R}. 
\begin{table}
\centering
\begin{tabular}{ccccc}
\hline
\hline
    Foreground              & $A[{\rm mK^2}]$ & $\beta$   &$\alpha$    &$\xi$   \\ \hline
    Galactic synchrotron    & 700       & 2.4       & 2.80       & 4.0 \\
    Point sources           & 57        & 1.1       & 2.07       & 1.0 \\
    Galactic free-free      & 0.088     & 3.0       & 2.15       & 35  \\
    Extragalactic free-free & 0.014     & 1.0       & 2.10       & 35  \\ \hline
\end{tabular}
\caption{Model parameters for the fiducial foreground $C_{\ell}\left(\nu_1, \nu_2\right)$ in Eq.~(\ref{eq:powerspectrum}), adapted from~\citep{2005ApJ...625..575S}, using the reference values of $\ell_{\text{ref}}=1000$ and  $\nu_{\text{ref}}=130$ MHz.}
\label{tab:santos}   
\end{table}

We also consider the presence of polarized foregrounds, where frequency-dependent Faraday rotation affects synchrotron emission, as discussed in references such as~\citep{1986rpa..book.....R}. This phenomenon leads to leakage into the unpolarized signal, which can potentially be a problematic contribution for foreground removal. In {\tt CRIME}, the influence of polarized synchrotron emission on foregrounds is characterized by two parameters: the Faraday-space correlation length ($\xi_\psi$) and the polarization leakage fraction ($\epsilon_\mathrm{p}$). Further details can be found in~\citep{2014MNRAS.444.3183A}.

Following the approach in {\tt CRIME}, we adopt a typical value of $\xi_\psi = 0.5~\mathrm{rad}~\mathrm{m}^{-2}$ to match simulations obtained using the {\tt Hammurabi} code \citep{2009A&A...495..697W}. For simplicity, we choose $\epsilon_\mathrm{p}=0.01$, which, importantly, depends entirely on the level of systematic control within the instrument. 

In the left panel of Fig.~\ref{fig:ClT_data}, we display the comparison of the angular power spectra. For these calculations, we employed the publicly available code \texttt{NaMaster}\footnote{https://github.com/LSSTDESC/NaMaster} \citep{NaMaster}, which was used to calculate the spectra among different components at a frequency of $\nu=1110$ MHz. Black solid lines are plotted for the cosmological HI, while other colored dashed or dotted lines represent different foreground components. It is evident that the foregrounds dominate the HI signal by 4-5 orders of magnitude in terms of $\sqrt{C_\ell}$.

The right panel of Fig.~\ref{fig:ClT_data} shows a randomly selected sky pixel as an example to demonstrate the dependence of brightness temperature on frequency. The black solid and blue dashed lines correspond to the cosmological HI and total foreground, respectively. As expected, the foreground temperature is smoothly distributed, whereas the HI exhibits significantly more fluctuations in temperature.

Fig. \ref{fig:CorrMatrix} displays the respective frequency-frequency correlation matrices of the cosmological HI, the total unpolarized foreground, and the polarized leakage, providing statistical evidence regarding the frequency coherence scales of the different components. Notably, the frequency correlation of the unpolarized foreground is much stronger than that of the HI signal. However, for the polarized synchrotron, which arises due to frequency-dependent Faraday rotation, the correlation lengths are significantly reduced. This non-smoothed component substantially increases the presence of foreground residuals in foreground-cleaned maps~\citep{2014PhRvD..89l3002D}.

\begin{figure*}
    \centering
    \includegraphics[trim=0cm 0cm 0cm 0cm, clip=True,width=2.1\columnwidth]{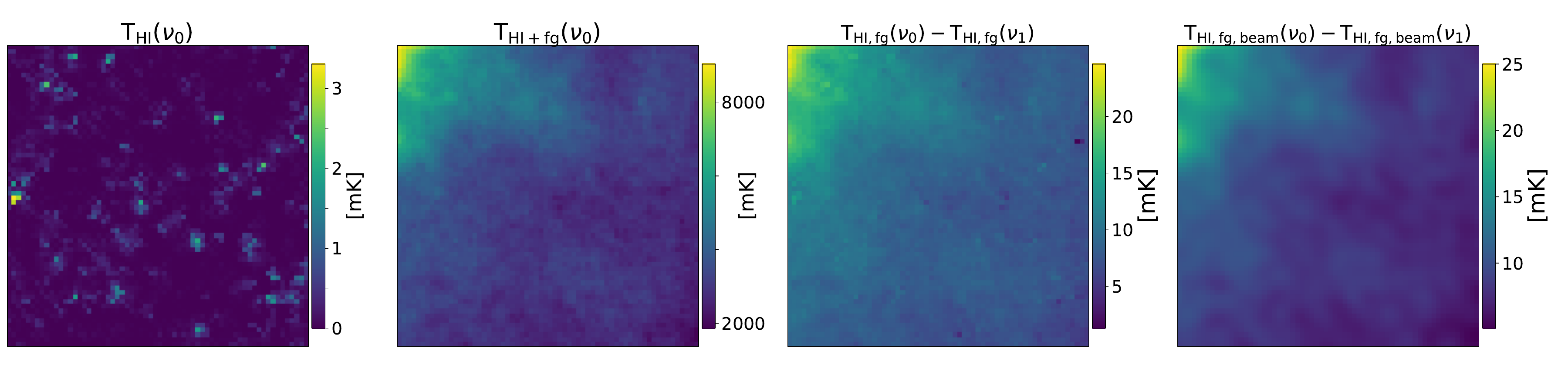}
    \caption{Visualization of the temperature difference between neighboring frequency bands. The first and second panels correspond to the pure HI temperature map $T_\mathrm{HI}(\nu_0)$ and the foreground added map $T_{\mathrm{HI},\mathrm{fg}}(\nu_0)$, at $\nu_0 = 1110$ MHz, respectively. The third panel displays the frequency-difference map $\Delta T = T_{\mathrm{HI},\mathrm{fg}}(\nu_0) - T_{\mathrm{HI},\mathrm{fg}}(\nu_1)$, where $\nu_1 = 1111$ MHz. The last panel shows the same frequency-difference map, but taking into account the beam effect, i.e., $\Delta T_\mathrm{beam} = B(\nu_0) \otimes T_{\mathrm{HI},\mathrm{fg}}(\nu_0) - B(\nu_1) \otimes T_{\mathrm{HI},\mathrm{fg}}(\nu_1)$.}
    \label{fig:Tms}
\end{figure*}

\subsection{Instrument effects}\label{sec:data_ns}

To simulate the effect of the beam of a radio telescope, we employ a Gaussian beam model for simplicity. This model approximately represents the primary beam response of a telescope as 
\begin{equation}\label{eq:beam}
    B(\nu,\theta) = {\rm exp}\left[-4{\ln}2\left(\frac{\theta}{\theta_{\rm FWHM}(\nu)}\right)^2\right]\,
\end{equation}
where $\theta_{\rm FWHM}$ represents the full width at half maximum (FWHM) of the primary beam. In Fig.~\ref{fig:bN_fre}, the blue solid line shows the beam size as a function of frequency across a total of 64 channels, spanning from 1100 MHz to 1164 MHz. Ideally, we would anticipate a smooth and gradual change in the beam size as the frequency varies. We then convolve our simulated sky maps (as described in previous sections) with such a beam model to incorporate the beam effect. However, in real observation, there would be frequency-dependent fluctuations in the beam size, which is called the ripple structure, e.g., \citep{2021MNRAS.506.5075M}. We plan to conduct a detailed analysis of this in a future study.

In the telescope receiver noise model, we solely consider uncorrelated white noise. We incorporate a Gaussian random field with a width of $\sigma_N$ into our simulated sky map, given by 
\begin{equation}
    \sigma_N=T_{\rm sys}\sqrt{\frac{4\pi f_{\rm sky}}
             {\Omega_{\rm beam}N_\mathrm{dish}t_\mathrm{obs}\Delta\nu}}\,,
\end{equation}
where $T_\mathrm{sys}$ is the total system temperature, which is the sum of the sky and receiver noise; $\Omega_{\rm beam} \approx 1.133\theta_{\rm FWHM}^2$ is the solid angle of the primary beam; $f_\mathrm{sky}$ is the fraction of sky coverage; $N_\mathrm{dish}$ denotes the number of dishes in the observation; $t_\mathrm{obs}$ indicates the integration time and $\Delta\nu$ is the bandwidth of each frequency channel.

In this study, to test the performance of our signal recovery method on thermal noise of varying amplitude, we rescale this white noise model by introducing a free parameter $\beta_{N}$, given by
\begin{equation}\label{eq:beta_ns}
    \beta_N = \frac{\sigma_N}{\sigma_\rmHI}\,,
\end{equation}
where $\sigma_\mathrm{HI}$ denotes the standard deviation of the HI signal over the simulation cube. Thus, $\beta_N$ is inversely proportional to the signal-to-noise ratio (S/N). In our analysis, we varied $\beta_N$ in the range of $0.0 < \beta_N < 1.0$, which is consistent with the noise amplitude desired from potential present and future intensity mapping configurations, e.g.~\citep{2021JCAP...04..081M}. In Fig.~\ref{fig:bN_fre}, the red dashed line represents the noise as a function of frequency, specifically for the case of $\beta_N=1$.

\section{Foreground removal method}\label{sec:method}
In this section, we will introduce our novel foreground removal method. By implementing the designed frequency-difference technique and utilizing the existing U-Net convolutional neural network (CNN)~\citep{2015arXiv150504597R}, we find that this strategy can efficiently recover cosmological HI signals with high fidelity by removing both smooth and non-smooth foregrounds.

\subsection{Frequency-difference technique}

\begin{figure*}[htpb]
    \centering
    \includegraphics[trim=0cm 0cm 0cm 0cm, clip=True,width=2.\columnwidth]{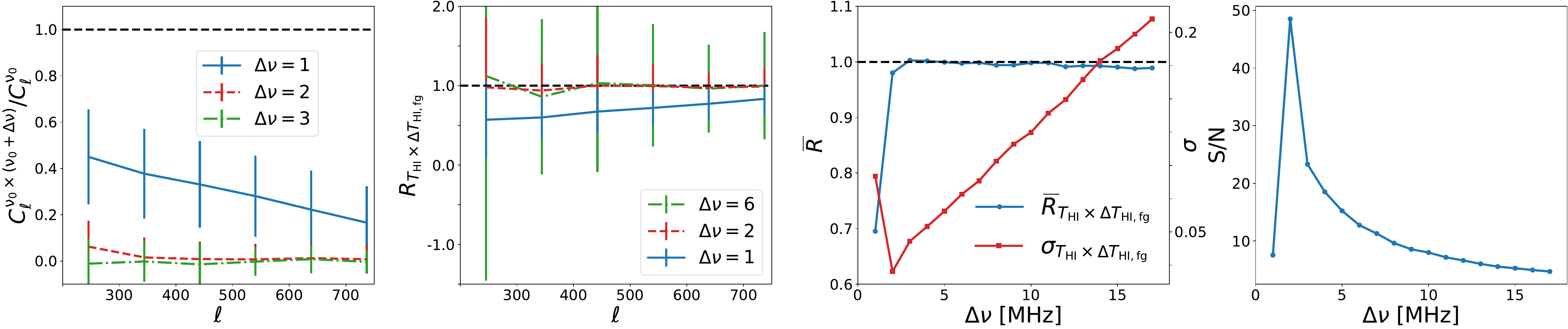}
    \caption{{\it First}: angular cross-correlation power spectrum between the pure HI at $\nu_0$ and that at $\nu_0+\Delta\nu$, $C^{\nu_0 \times (\nu_0+\Delta\nu)}$, which is normalized to the auto-correlation power spectrum of the pure HI at $\nu_0$. Different lines correspond to the results of different $\Delta \nu$, as indicated. {\it Second}:  ratios of the angular power spectra for different $\Delta \nu$, $R_{T_\rmHI \times \Delta T_{\rmHI,\rmfg}}=C^{T_\rmHI \times \Delta T_{\rmHI,\rmfg}}_\ell/C^{T_\rmHI}_\ell$, where $C^{T_\rmHI \times \Delta T_{\rmHI,\rmfg}}$ is the cross-correlation power spectrum between HI map, $T_\rmHI(\nu_0)$, and frequency-difference map, $\Delta T_{\rmHI,\rmfg} = T_{\rmHI, \rmfg}(\nu_0)-T_{\rmHI, \rmfg}(\nu_0 + \Delta\nu_0)$, and $C^{T_\rmHI}$ denotes the auto-correlation power spectrum of $T_\rmHI(\nu_0)$. The blue solid, red dashed, and green dashed-dot lines correspond to $\Delta \nu= 1$, $2$, and $6$ MHz, respectively. Here, the error bars correspond to the $1\sigma$ statistical uncertainty estimated from among the all 192 sky patches. {\it Third}: mean ratio (averaged over all sky patches for the $C_\ell$ values in the range of $\ell=$ 200 to 800), $\bar{R}_{T_{\mathrm{HI}} \times \Delta T_{\mathrm{HI},\mathrm{fg}}}$ (blue line, and its value is related to the left y-axis), and the RMS error, $\sigma_{T_\mathrm{HI} \times \Delta T_{\mathrm{HI},\mathrm{fg}}}$ (red line, and its value is related to the right y-axis), as a function of $\Delta\nu$. {\it Fourth}: S/N as a function of $\Delta \nu$ according to Eq.~(\ref{eq:SNR}).}
    \label{fig:Clms}
\end{figure*}

One of the key ingredients of our method is to reduce the dynamic range of foreground amplitudes before feeding them into U-Net. Based on the characterization of the frequency correlation discussed in Sec.\ref{sec:fgrd}, {\it we use the difference in the sky map temperature between two neighboring frequency bands as the input, which we call the frequency-difference technique}. In this case, the foreground temperatures of the neighboring bands can largely cancel each other out due to the strong correlation in frequency. Conversely, The HI signal can be preserved in the frequency-difference map because the HI signal becomes uncorrelated between adjacent frequency bands when the bandwidth $\Delta \nu \gtrsim 2$ MHz (as indicted in Fig.~\ref{fig:Clms}). 

Fig.~\ref{fig:Tms} presents a visual comparison of temperature maps and their differences between neighboring frequency bands. The first and second panel correspond to the pure-HI temperature map, $T_\rmHI(\nu_0)$, and the HI plus foreground map, $T_{\rmHI, \rmfg}(\nu_0)$ respectively, where $\nu_0=1110~\rmMHz$. The third one shows the temperature difference between neighboring bands, $T_{\rmHI, \rmfg}(\nu_0)-T_{\rmHI, \rmfg}(\nu_1)$, where $\nu_1=1111$ MHz. It is clear that due to the strong correlation of the foreground in frequency, the amplitude range is significantly reduced by $2-3$ orders of magnitude by means of this frequency-difference technique. 
Furthermore, the last panel shows the frequency-difference map while considering the beam. In this regard, it is important to consider that the size of the beam ($\theta_\mathrm{FWHM}$) varies with frequency. To ensure consistency in the beam size level, we have smoothed each adjacent-band map to match their relatively lower angular resolution. This was done by convolving the higher frequency map with a Gaussian beam characterized by a beam size of $\Delta\theta_\mathrm{FWHM}$, denoted as,
\begin{equation}
    \Delta\theta_\mathrm{FWHM} = \sqrt{\theta_\mathrm{FWHM,low}^2-\theta_\mathrm{FWHM,high}^2}.
\end{equation}
Here $\theta_\mathrm{FWHM, low}$ and $\theta_\mathrm{FWHM, high}$ correspond to the beam size of low and high frequency, respectively. It reveals that the frequency-difference map yields a similar outcome, with a significant reduction in the range of amplitude.
% the amplitude range is also suppressed dramatically in the frequency-difference map after beam convolution.}

% exhibits similar results  dramatically suppresses the dynamic range
% of the foreground amplitudes.}
%Therefore, the difference map retains foreground contamination at the small scales from the frequency map with the smallest beam $\theta_\mathrm{FWHM}$ value.} 

%}

In the left panel of Fig.~\ref{fig:Clms}, we show the angular cross-correlation power spectrum, $C^{\nu_0 \times (\nu_0+\Delta\nu)}$ between the pure HI at $\nu_0$ and that at $\nu_0+\Delta\nu$, which is normalized to the auto-correlation power spectrum of the pure HI at $\nu_0$. Different lines correspond to the results of different $\Delta \nu$, as indicated. We can see a clear positive correlation signal at $\Delta\nu=1$. This is due to the fact that this bandwidth corresponds to $\Delta z \simeq 0.0012$, and hence the comoving distance is $\Delta r \simeq 4.5 \mpch$. Thus, the presence of HI large-scale structures (e.g., filaments, voids, and sheets) allows the maps between the two frequencies to remain correlated. Therefore, temperature differences between neighboring bands ($\Delta\nu=1$) might also lead to a loss of the HI signal. However, when $\Delta\nu \geq 2$ MHz, there is no longer a correlation of the HI signals between the sky maps, resulting in the corresponding cross-correlation signals decaying rapidly to zero.

In the second panel of Fig.~\ref{fig:Clms}, which aims to demonstrate the correlation of the frequency-difference map with the pure HI signal for different $\Delta \nu$ values, we present the ratios obtained by varying $\Delta \nu$. Here, the ratio is defined as $R_{T_\rmHI \times \Delta T_{\rmHI,\rmfg}}=C^{T_\rmHI \times \Delta T_{\rmHI,\rmfg}}_\ell/C^{T_\rmHI}_\ell$, where $C_\ell^{T_\rmHI \times \Delta T_{\rmHI,\rmfg}}$ is the angular cross-correlation power spectra between $T_\rmHI(\nu_0)$ and $\Delta T_{\rmHI,\rmfg} = T_{\rmHI, \rmfg}(\nu_0)-T_{\rmHI, \rmfg}(\nu_0 + \Delta\nu_0)$, and $C_\ell^{T_\rmHI}$ is the angular auto-correlation power spectra of $T_\rmHI(\nu_0)$. The blue solid, red dashed, and green dashed-dot lines correspond to $\Delta \nu= 1$, 2, and 6 MHz, respectively. The error bars represent the standard deviation estimated across the 192 sky patches, as mentioned in Sec.\ref{sec:data}. 

In the cases of $\Delta \nu = 2$ \& $6$ MHz, the ratios are positive and close to unity, proving that the frequency-difference helps to reduce the foreground contamination and keep the HI signal visible. 

When $\Delta \nu = 1$ MHz, however, the ratios are slightly underestimated. This indicates a loss of the HI signal due to large-scale HI correlations\footnote{The HI correlation is expected to be enhanced by the small-scale Redshift Space Distortion (RSD) effect (the so-called Finger of God effect.). However, in this paper, we do not consider the impact of this effect when using the linear velocity simulations in {\tt CRIME}}. On the other hand, the error bars for $\Delta \nu = 6$ are considerably larger than those for $\Delta \nu = 2$. This is due to the reduced correlation between the foreground sky maps at wider frequency intervals, leading to more substantial residual foregrounds on the frequency-difference maps.

In the third panel of Fig.~\ref{fig:Clms}, by varying $\Delta\nu$ from 2 to 17 MHz, we present the mean ratios averaged over all sky patches, $\bar{R}_{T_\rmHI \times \Delta T_{\rmHI,\rmfg}}$ (blue line), and the RMS error, $\sigma_{T_\rmHI \times \Delta T_{\rmHI,\rmfg}}$ (red line).  
The results are computed based on $C_\ell$ values within the range of $\ell=200$ to 800. The ratios quickly converge to unity when $\Delta \nu > 1$ MHz. However, the uncertainties first reach their minimum at $\Delta \nu = 2$ MHz and then rapidly increase beyond that.  Therefore, the width $\Delta \nu$ needs to be optimized to minimize the HI correlation while maximizing the foreground correlation between two frequency bands. To do so, we then compute the signal-to-noise ratios (S/N) as a function of $\Delta\nu$ by,
\begin{equation}\label{eq:SNR}
    \mathrm{S/N}\left(\Delta\nu\right) = \sqrt{\frac{\bar{R}_{T_\rmHI \times \Delta T_{\rmHI,\rmfg}}}{\sigma_{T_\rmHI \times \Delta T_{\rmHI,\rmfg}}}}.
\end{equation}
The S/N result are depicted on the right panel of Fig.~\ref{fig:Clms}. When $\Delta\nu=2$,  a peak occurs, indicating the potential increase in sensitivity for detecting HI signals.

So, the frequency-difference technique has the potential to significantly reduce foreground contamination. By selecting an appropriate bandwidth, it becomes possible to efficiently diminish foreground residuals in the resulting maps while enhancing the S/N ratio. Additionally, when considering the effect of beam smoothing, the frequency-difference maps at small-scales are smeared out (as seen in the last panel of Fig.~\ref{fig:Tms}), which can lead to a challenge in recovering the HI signal.
\begin{figure*}
    \centering
    \includegraphics[width=2\columnwidth]{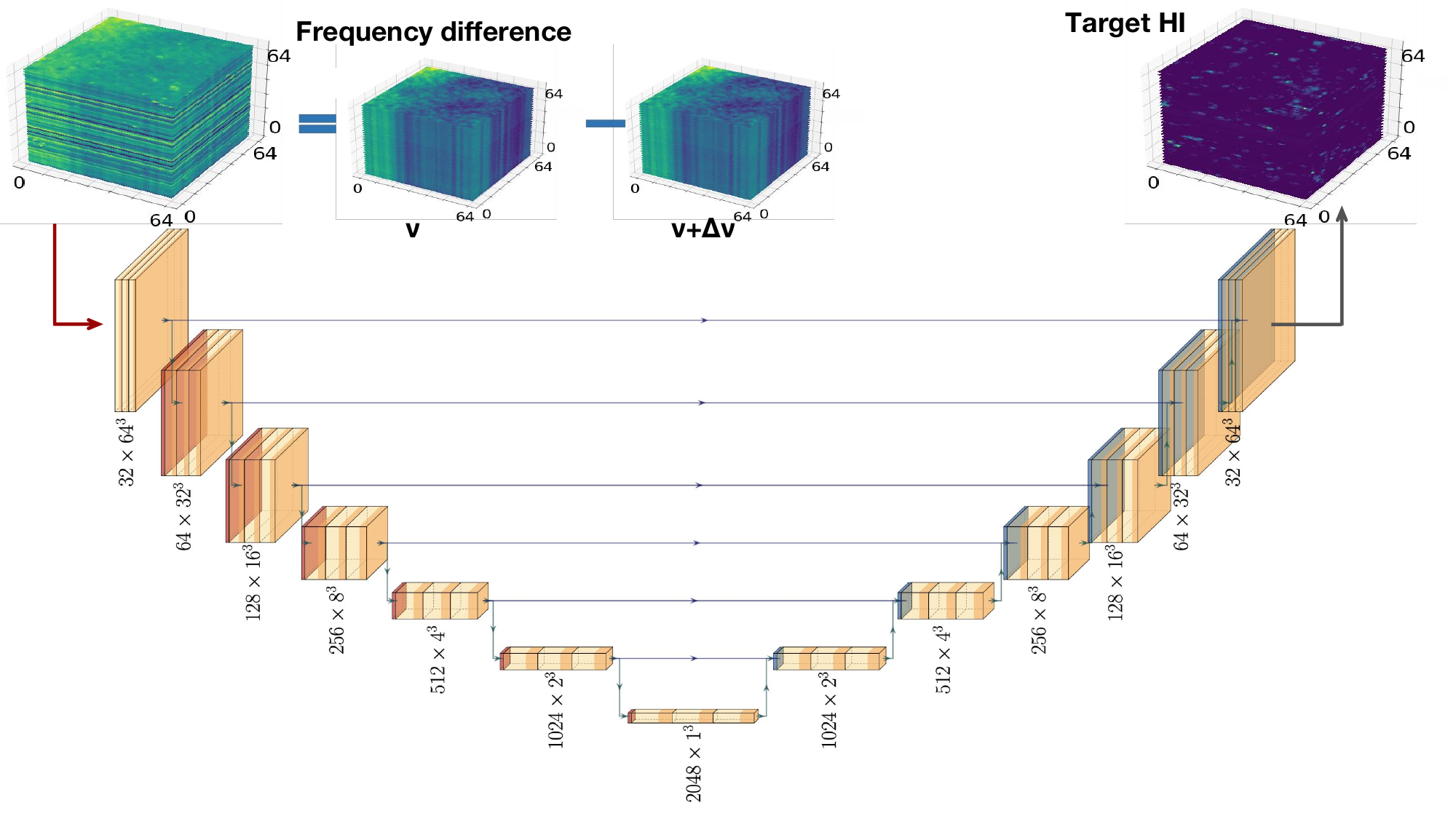}
    \caption{Visualization of the UNet-fd architecture. The input and output cubes, each containing grids of size $64^3$, correspond to the frequency-difference maps, where the difference between two frequencies is denoted as $\Delta \nu$, and the target HI temperature maps, respectively. The architecture comprises a total of 13 layers, depicted by the orange box. These layers consist of 6 in the encoder and decoder each, with an additional layer for the bottleneck. Each layer consists of 3 convolutional operations, incorporating batch normalization and activation functions (except for the final convolutional block on the output side). The dark red box and the dark blue box indicate the input and output of each layer. The bottom of each box illustrates the number of channels and the size of the output. The down, right, and up arrows indicate maximum pooling, skip connection, and transpose convolutions, respectively. This visualization is made with the \tt{PlotNeuralNet} library.} 
    \label{fig:Unet_arch}
\end{figure*}
\begin{figure*}
    \centering
    \includegraphics[trim=0cm 0cm 0cm 0cm, clip=True,width=1.7\columnwidth]{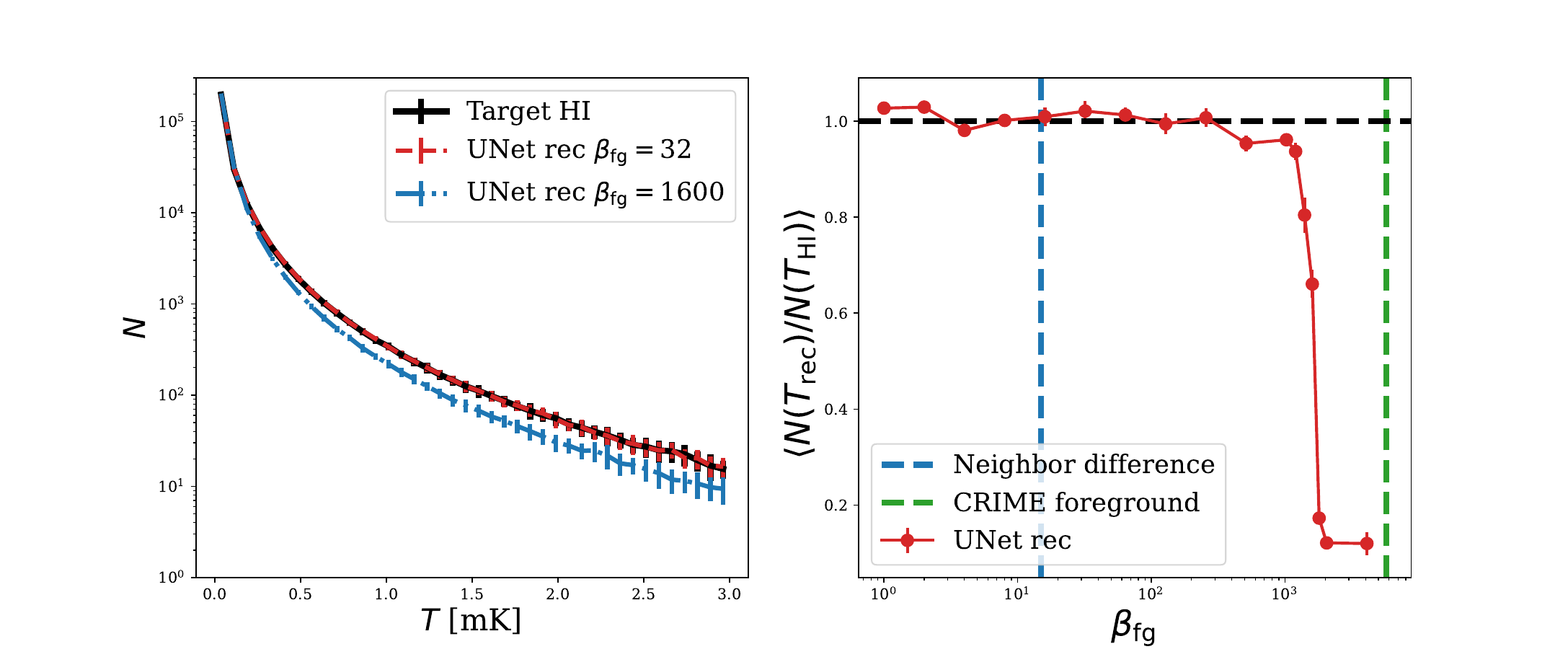}
    \caption{Testing UNet-fd cleaning in different-level foreground cases. {\it Left}: comparisons of temperature distribution as a function of $T$. The back solid line represents the target HI. The red dashed (blue dash-dotted) line is shown for the U-Net-reconstructed results of $\beta_\mathrm{fg} = 32$ ($\beta_\mathrm{fg}=1600$), where $\beta_\mathrm{fg}$ is the foreground strength parameter as defined in Eq.~(\ref{eq:beta_fg}). {\it Right:} mean ratios of temperature distribution, $\left<N(T_\mathrm{rec})/N(T_\mathrm{HI})\right>$, between the U-Net-reconstructed $N(T_\mathrm{rec})$ and the target $N(T_\mathrm{HI})$. The red dotted solid line represents the U-Net-reconstructed results as a function of $\beta_\mathrm{fg}$. The blue and green vertical dashed lines are shown for the $\beta_\mathrm{fg}$ level of the frequency-difference ($\Delta \nu =2$) case and the CRIME-simulated foreground case (which is the original output foreground of CRIME as described in Sec.~\ref{sec:fgrd}), respectively.}
    \label{fig:Pkr_fglevel}
\end{figure*}

\begin{figure*}
    \centering
    \includegraphics[trim=4cm 2cm 4cm 2cm, clip=True,width=1.9\columnwidth]{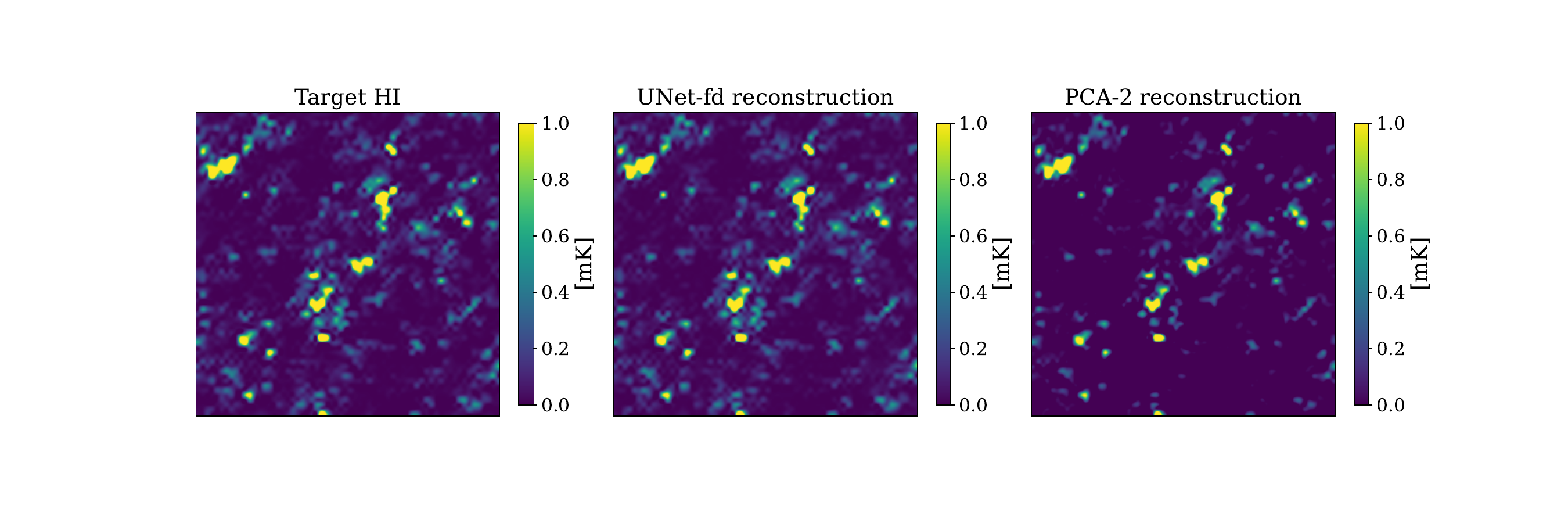}
    \caption{Comparison of UNet-fd and PCA reconstructed HI maps for Case I. From left to right, we show the target HI, the UNet-fd reconstruction, and the PCA reconstruction, respectively. The data cube is randomly selected from the test set, and the slice shown here corresponds to a frequency of $\nu=1130$ MHz.}
    \label{fig:img_Hf}
\end{figure*}
\begin{figure*}
    \centering
    \includegraphics[trim=0cm 0cm 0cm 0cm, clip=True,width=2\columnwidth]{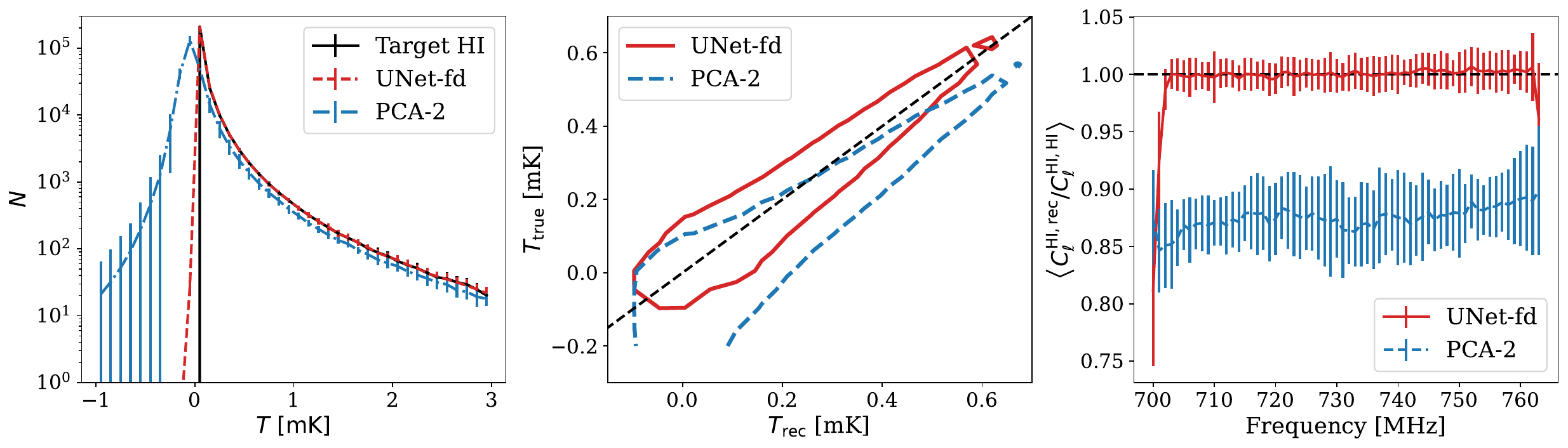}
    \caption{The results for Case I. {\it Left}: the comparison of the temperature distributions among the truth (black),  PCA (blue) and UNet-fd (red). {\it Middle}: the reconstructed $T_\rmrec$ versus the true $T_\rmtrue$. The results from UNet-fd and PCA are respectively shown with red solid and blue dashed contours, which include about 95\% grid cells. {\it Right}: the mean ratio of angular power spectrum between $C^{\rmHI,\rmrec}_\ell$ and $C^{\rmHI,\rmHI}_\ell$ as a function of the frequency channel. Red solid and blue dashed lines correspond to the results of UNet-fd and PCA, respectively. Error bars correspond to 1$\sigma$ statistical uncertainty, estimated from the test sets. }
    \label{fig:HPR_Hf}
\end{figure*}

\subsection{U-Net Network}

To further mitigate foreground contamination while recovering the HI signal, we employ a deep neural network based on the U-Net architecture~\citep{2015arXiv150504597R}. We will demonstrate that the combination of the frequency-difference technique and U-Net can yield an outperformed performance in recovering HI signals. U-Net is a versatile model for image-to-image translation, which incorporates structural modifications into the Convolutional Neural Network (CNN) framework. These modifications enable U-Net to operate effectively even with limited training images, resulting in more accurate translations.

Fig.~\ref{fig:Unet_arch} provides a visualization of our model architecture, which we refer to as ``UNet-fd" (UNet frequency-difference). The input and output cubes, both containing $64^3$ grids, represent the frequency-difference maps and the reconstructed HI temperature maps, respectively. The U-Net maps the input cube to the output cube through a symmetric encoder-decoder convolution scheme. The encoder part, which is comprised of 6 convolutional layers, downsamples the input to the bottleneck, where the spatial information is reduced and feature information is increased. Skip connections are used to connect layers in the encoder with corresponding layers in the decoder, allowing to preserve the spatial information during downsampling. The decoder part then upsamples the bottleneck to the required output size by using 6 transposed convolutional layers. Following~\cite{2021JCAP...04..081M}, we perform 3 convolutions at each layer, and apply batch normalization and ReLU activation between convolutional layers, except for the final convolutional block on the output side.

To train the networks, we use the Logcosh loss function,
\begin{equation}
        \mathcal{L}(T_\mathrm{p},T_\mathrm{t}) = \mathbb{E}\left[ \mathrm{log~cosh}(T_\mathrm{p}-T_\mathrm{t})\right],
\end{equation}\label{eq:loss}
where $T_\mathrm{p}$ denotes the predicted temperature and $T_\mathrm{t}$ denotes target truth. The Logcosh loss function has been tested to be more robust and less sensitive to outliers \citep{2021JCAP...04..081M}. We train the network from scratch and use the {\tt AdamW} \citep{2017arXiv171105101L} optimizer, which provides weight decay regularization. The hyperparameters are fine-tuned to optimize the network, following the configuration of {\tt Deep21} \citep{2021JCAP...04..081M}. The network processes 140 training samples each epoch, followed by 12 validation samples used solely for model evaluation during training. Finally, we save our best model based on the minimum validation-set loss function.

In the training phase, the UNet model is tasked with extracting the targeted HI signal at a given frequency when provided with the frequency-difference map as input. This map combines the signals from two neighbouring frequency bands into a single channel. To understand how the UNet model process in extracting the targeted single-band HI from the difference map, a detailed explanation is provided in the Appendix.

On the other hand, we found that without the frequency-difference preprocessing, directly using the U-Net performs poorly in recovering the HI signal. To demonstrate the threshold limit for foreground cleaning by the U-Net, we ran the network by feeding the HI cubes with different-level foregrounds added. To do so, we vary the CRIME-simulated foreground temperature (which is the original output foreground of CRIME as described in Sec.~\ref{sec:fgrd}) to the desired amplitude level, which is parametrized by
\begin{equation}\label{eq:beta_fg}
    \beta_\mathrm{fg} = \sqrt{\frac{\left<(T_\rmfg-\overline{T}_\rmfg)^2\right>}
                 {\left<(T_\rmHI-\overline{T}_\rmHI)^2\right>}}\,,
\end{equation}
representing the square root of the ratio between the foreground temperature variance and the HI signal variance averaged over pixels.

We then rescale the foreground to the 17 levels of $\beta_\mathrm{fg}: 1$, $2$, $4$, $8$, $16$, $32$, $64$, $128$, $256$, $512$, $1024$, $2048$, and $4096$. By adding the rescaled foregrounds on the pure HI map, we run U-Net repeatedly for each of the 17 cases and obtain the best-fitting models based on the training samples described above. We then apply the resulted models on the 40 test samples and compare the recovered temperatures with the target HI. In the left panel of Fig.~\ref{fig:Pkr_fglevel}, we show the temperature distribution as a function of $T$. The back solid line represents the target HI. The red dashed and blue dash-dotted lines are shown for the U-Net-reconstructed results of $\beta_\mathrm{fg}=32$ and $\beta_\mathrm{fg}=1600$, respectively. The results are averaged over the test samples. As can be seen, the target truth and the reconstructed result of $\beta_\mathrm{fg}=32$ coincide nicely, however, the result of $\beta_\mathrm{fg}=1600$ is systematically underestimated. Fig.~\ref{fig:Pkr_fglevel} shows the mean ratios of temperature distribution, $\left<N(T_\mathrm{rec})/N(T_\mathrm{HI})\right>$, between the U-Net-reconstructed $N(T_\mathrm{rec})$ and the target $N(T_\mathrm{HI})$. The red dotted solid line represents the U-Net-reconstructed results as a function of $\beta_\mathrm{fg}$. The blue and green vertical dashed lines are shown for the $\beta_\mathrm{fg}$ level of the frequency-difference ($\Delta \nu =2$) case and the CRIME-simulated foreground case respectively, where the former has $\beta_\mathrm{fg} \sim 30$ and the latter has $\beta_\mathrm{fg} \sim 5700$. It is evident that the ratios keep values consistent to unity for $\beta_\mathrm{fg}<100$ and 10\% reduction at $\beta_\mathrm{fg}=1000$; however, for $\beta_\mathrm{fg} > 1000$, the ratios systematically decreases. Therefore, it is virtually impossible to remove the CRIME-simulated foreground contamination using U-Net alone. This is primarily due to the large dynamic range of the map amplitude \citep{2002ApJ...564..576D, 2008MNRAS.388..247D,2014PhRvD..90b3018L}, which makes extracting correct information with the U-Net network difficult \citep{2021JCAP...04..081M, 2022ApJ...934...83N}. Fortunately, the frequency-difference $\beta_\mathrm{fg}$ value falls within the U-Net's working range. We then expect to have a recovered HI temperature that was close to the target truth by applying the U-Net on the frequency-difference map. 

\begin{figure*}
    \centering
    \includegraphics[trim=0cm 0cm 0cm 0cm, clip=True,width=1.5\columnwidth]{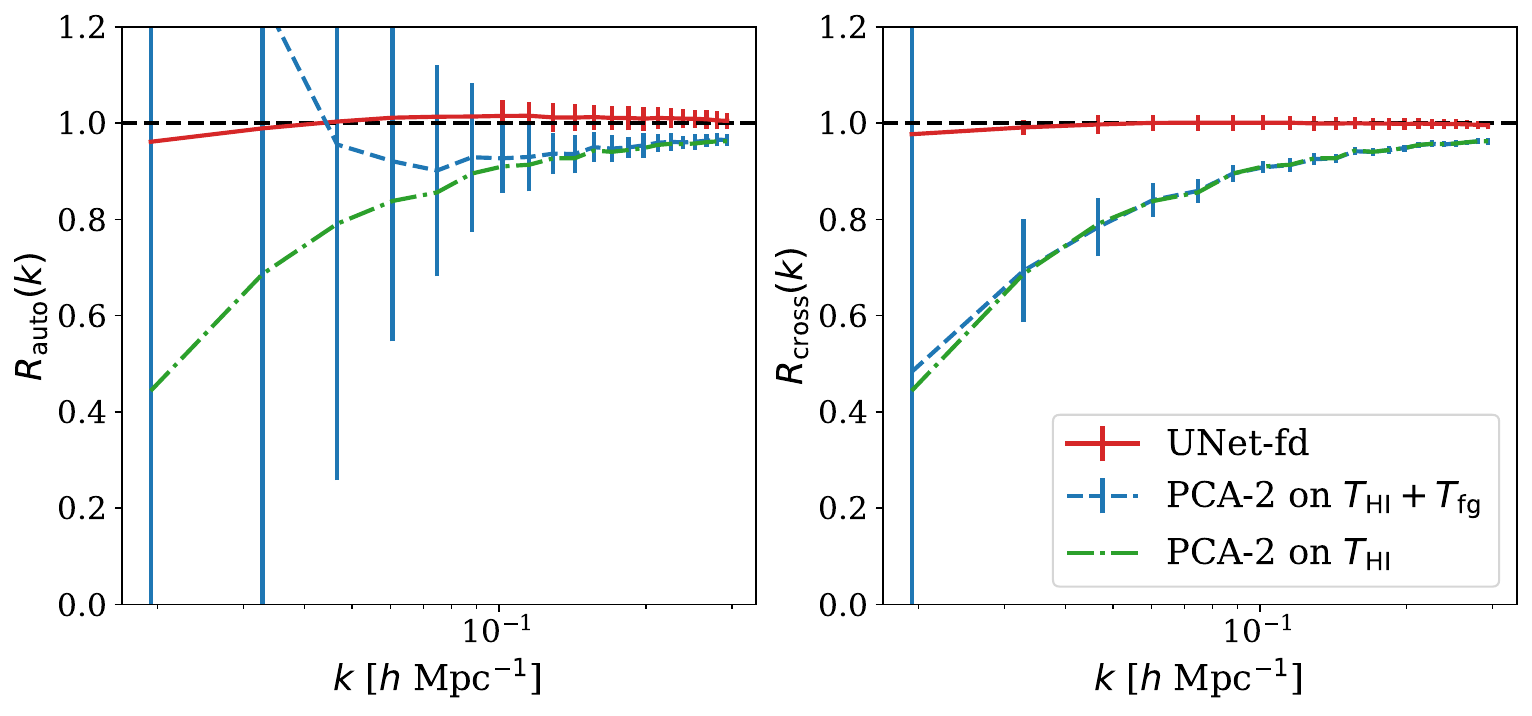}
    \caption{For Case I, ratios of auto-correlation (left panel) and cross-correlation (right panel) $P(k)$ between the reconstruction and the true HI as defined in Eqs.~\ref{eq:Rauto} and \ref{eq:Rcross}. Red solid and blue dashed lines correspond to UNet-fd and PCA results reconstructed from the map of $T_\rmHI+T_\rmfg$. Green dash-dotted lines show the results obtained by subtracting PCA modes on the pure HI map.  Error bars indicate the $\pm$ sigma variance among the 40 test sets.}
    \label{fig:Pk1d_Hf}
\end{figure*}
\begin{figure*}
    \centering
    \includegraphics[trim=0cm 2cm 0cm 3cm, clip=True,width=1.6\columnwidth]{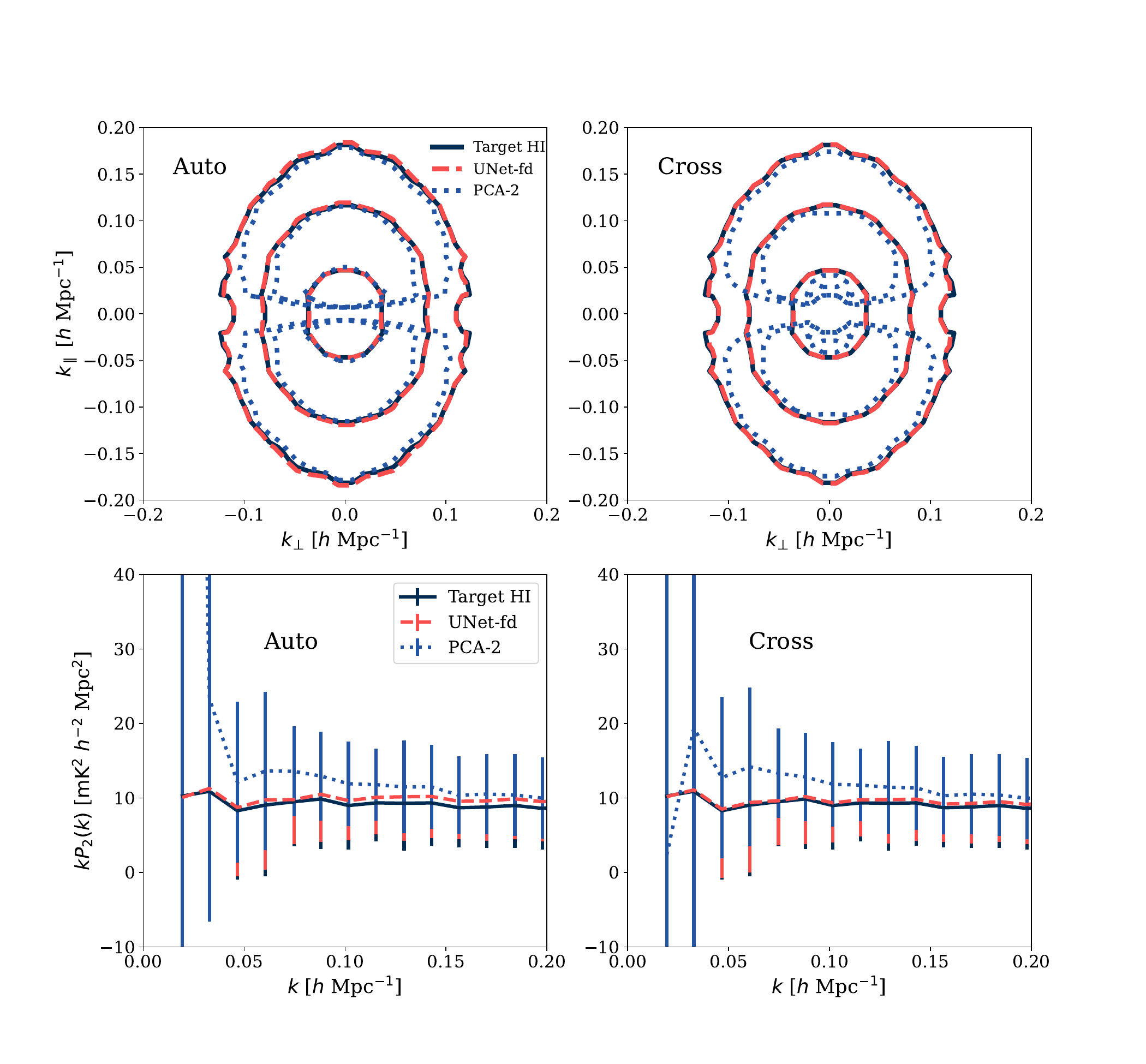}
    \caption{Same as in Fig.~\ref{fig:Pk1d_Hf}, but for the 2D power spectra for Case I. {\it Upper}: comparison of $P(k_\bot, k_\parallel)$, where black solid, red dashed, and blue dotted contours correspond to the true HI, the UNet-fd reconstruction, and the PCA reconstruction, respectively. The contour level corresponds to $P(k) = 900$, 300, and 180 $h^{-3} \mathrm{Mpc^3}$. The results are averaged from the 40 test sets. {\it Lower}: comparison of $P_2(k)$. Black solid, red dashed, and blue dotted contours correspond to the true HI, the UNet-fd reconstruction, and the PCA reconstruction, respectively. Error bars indicate the $\pm1$ variance among the 40 test sets. The lower-left and lower-right panels correspond to auto-correlation and cross-correlation power spectrum, respectively. }
    \label{fig:Pk2d_Hf}
\end{figure*}

\section{Results}\label{sec:result}
In this section, we present our results of the foreground removal. In order to verify the impact of various contaminations, we implement the tests considering the following three cases:
\begin{itemize}
\item Case I: add the cosmological HI signal and the foregrounds without the presence of the beam effect and thermal noise. In other words, the observed data ($d$) is simply the sum of the foreground ($f$) and HI signal ($s$), such that $d = s + f$.
\item Case II: in comparison to Case I, we further consider the beam ($A$) convolution effect, which can be expressed as $d = A \otimes (s + f)$.
\item Case III: to mimic a realistic observation, we introduce white noise ($n$) into the data. In this case, the data can be represented as $d = A \otimes (s + f) + n$.
\end{itemize}
For each case, we run the U-Net training independently. We also compare our results with those from the PCA method. In our analysis, we subtract the first two and three modes for Case I and Case II, separately. Hence, the notation PCA-$N$ corresponds to the reconstruction for which the first $N$ principal components have been removed.

\subsection{Case I: Adding HI and foreground}

We start with the test for Case I. Fig.~\ref{fig:img_Hf} shows the comparisons in a {\tt HEALPix} pixel slice randomly drawn from the test set at frequency channel $\nu=1130$ MHz. Different panels correspond to the target HI, the UNet-fd reconstruction, and the PCA-2 reconstruction, as indicated at the top of each panel. The UNet-fd reconstruction exhibits recognizable structures that are in excellent agreement with the target signals over a range of temperature regions. In contrast, PCA underperforms and over-subtracts the signal, especially in low-temperature regions.

Fig.~\ref{fig:HPR_Hf} exhibits quantitative comparisons. The left panel shows the temperature distributions. The error bars reflect the $\pm 1 \sigma$ variance among the 40 test slices. Black solid, red dashed and blue dash-dotted lines correspond to the results of the target, UNet-fd, and PCA-2, respectively. As we can see, the UNet-fd reconstruction consistently agrees with the target, but the PCA reconstruction shows a systematical underestimation and reproduces nearly half of the negative-temperature values due to over-subtraction. The middle panel shows the temperature relation between the reconstructed and the target grid cells. The results from UNet-fd and PCA-2 are respectively shown with red solid and blue dashed contours, which encompass about 95\% grid cells. The UNet-fd reconstruction is linearly correlated with the true field and does not exhibit any significant bias, while the PCA result shows a highly biased relation.

To compare the reconstructed HI distribution at each frequency, we estimate the ratio of the angular power spectra between the cross-correlation, $C^{\rmHI,\rmrec}_\ell$, and the auto-correlation, $C^{\rmHI,\rmHI}_\ell$ as a function of the frequency channel in the right panel of Fig.~\ref{fig:HPR_Hf}. Red solid and blue dashed lines correspond to the results of UNet-fd and PCA-2 respectively, which are averaged from $C_\ell$ within the range of $0<\ell<760$. Error bars indicate the $\pm 1 \sigma$ variance among the test sets. The UNet-fd ratios agree very well with the unity over middle-frequency channels, proving that the HI signal was successfully recovered overall. 
The cross-correlations are underestimated at the boundary-frequency channels, showing the existence of the boundary effect in our method. As a comparison, the PCA results have cross-correlations with about 13\% reduction over all frequencies.

To further verify the clustering of the reconstructed signal in 3D space, we measure the power spectra $P(k)$, 2D power spectra $P(k_\bot, k_\parallel)$, and quadrupole $P_2(k)$ in the cubes of test sets, where we used the {\tt pylians} \footnote{https://github.com/franciscovillaescusa/Pylians3/blob/masterz
/docs/source/citation.rst} code \citep{Pylians} for calculating. Fig.~\ref{fig:Pk1d_Hf} shows the ratios of auto- and cross-correlation $P(k)$ between the reconstruction and the target HI, which are respectively defined as,  
\begin{equation}\label{eq:Rauto}
            R_\mathrm{auto}(k) = \frac{P_{\rmrec,\rmrec}(k)}{P_{\rmHI,\rmHI}(k) },\\
\end{equation}
and,
\begin{equation}\label{eq:Rcross}
            R_\mathrm{cross}(k) = \frac{P_{\rmrec,\rmHI}(k)}{P_{\rmHI,\rmHI}(k)}.
\end{equation}
Here $P_{\rmHI,\rmHI}(k)$ and $P_{\rmrec,\rmrec}(k)$ are the auto-correlation power spectrum of the target HI and the reconstructed HI respectively, and $P_{\rmrec,\rmHI}(k)$ is their cross-correlation power spectrum. Red solid and blue dashed lines correspond to the UNet-fd and the PCA-2 reconstructions, respectively. Error bars indicate the 1-$\sigma$ uncertainty among the 40 test sets. Both auto-correlation and cross-correlation ratios of UNet-fd agree excellently with unity over the ranges of scales $k<0.3\hmpc$, indicating a successful recovery of the HI clustering and remarkable concordance between the reconstructed and the true HI temperature field. It is significant that UNet-fd
outperforms PCA, which, especially at large scale, suffers from residual foregrounds and signal loss. On the other hand, the method of \citet{2021JCAP...04..081M} uses PCA to subtract a few modes before feeding to the UNet. Although the method can potentially improve performance, the reduced HI signal that is eliminated by PCA is virtually impossible to be recovered. To show the level of signal loss subtracted by PCA, the green dash-dotted lines show the results obtained by subtracting the first two modes on the pure HI map. There are both underestimations of $R_\mathrm{auto}(k)$ and $R_\mathrm{cross}(k)$ due to signal loss, which is unexpected to be avoided by the PCA-UNet strategy. As a comparison, it is reassuring that our UNet-fd method is immune to the signal loss issue.

\begin{figure*}
    \centering
    \includegraphics[trim=4cm 0cm 2cm 0cm, clip=True,width=2\columnwidth]{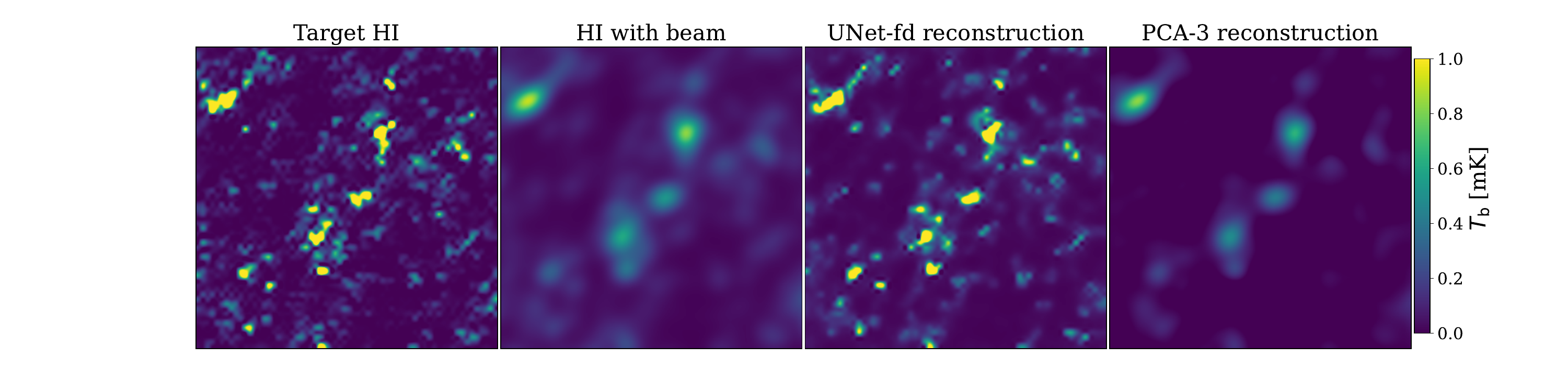}
    \caption{Comparison of the brightness temperature field for Case II (considering the beam effect). From left to right, we show the target HI, the target HI convolved with the beam (according to Eq.~(\ref{eq:beam})), the UNet-fd reconstruction, and the PCA reconstruction, respectively. The target HI is selected to be same as the one in Fig.~\ref{fig:img_Hf}. As shown, our neural network not only removes strong foregrounds but also automatically performs beam deconvolution with high accuracy.} \label{fig:img_Hfb}
\end{figure*}
\begin{figure*}
    \centering
    \includegraphics[trim=0cm 0cm 0cm 0cm, clip=True,width=2.0\columnwidth]{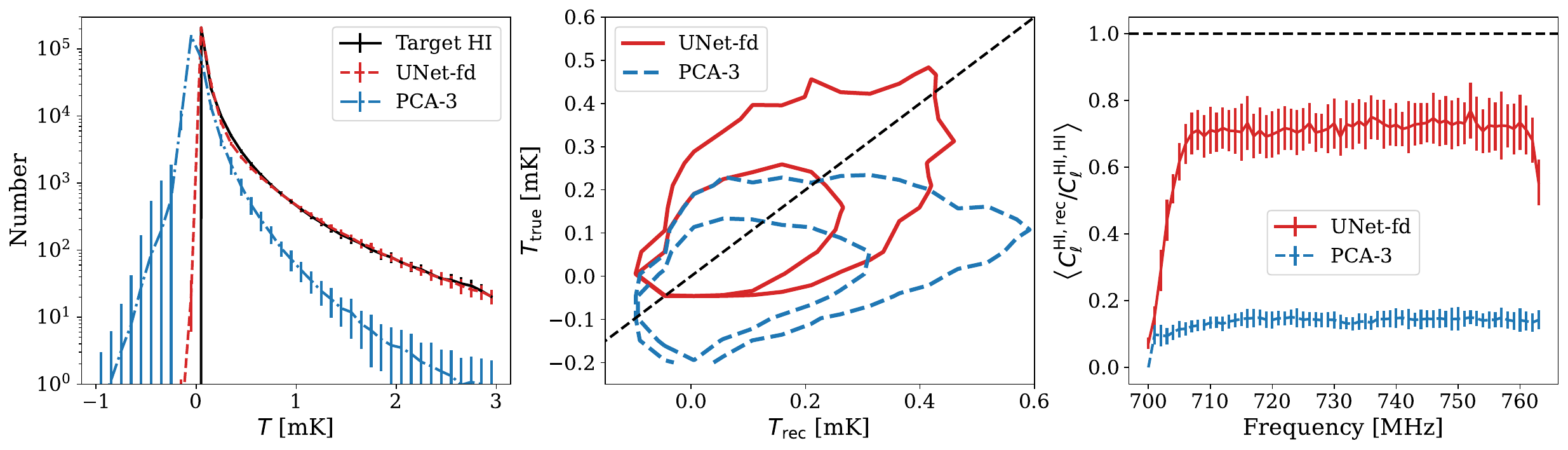}
    \caption{Same as in Fig.~\ref{fig:HPR_Hf}, but for testing the beam effect for Case II. In the middle panel, the two contours of each case encompass 90\% and 95\% of the grid cells, respectively.}
    \label{fig:HPR_Hfb}
\end{figure*}
\begin{figure*}
    \centering
    \includegraphics[trim=0cm 0cm 0cm 0cm, clip=True,width=1.5\columnwidth]{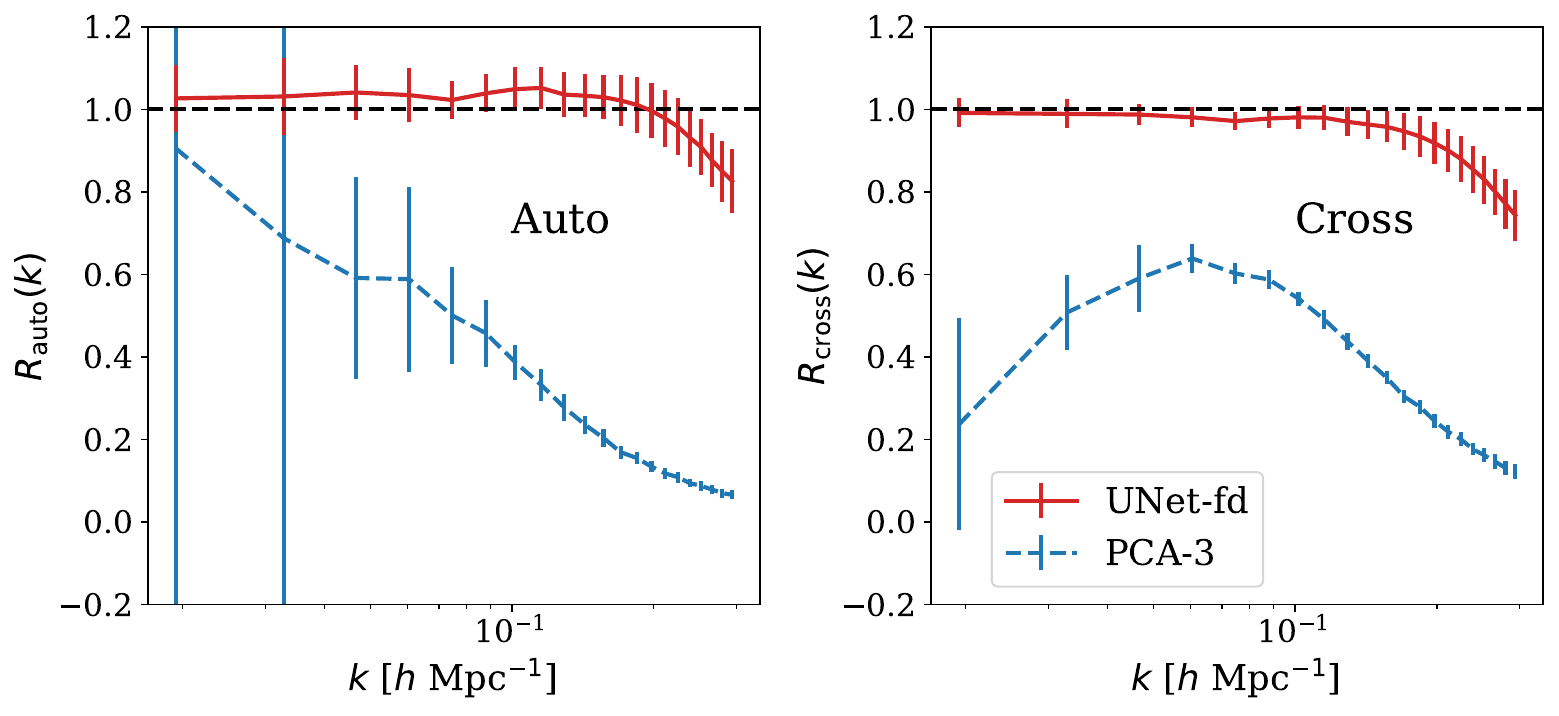}
    \caption{Same as in Fig.~\ref{fig:Pk1d_Hf}, but for Case II.}
    \label{fig:Pk1d_Hfb}
\end{figure*}
\begin{figure*}
    \centering
    \includegraphics[trim=0cm 2cm 0cm 3cm, clip=True,width=1.6\columnwidth]{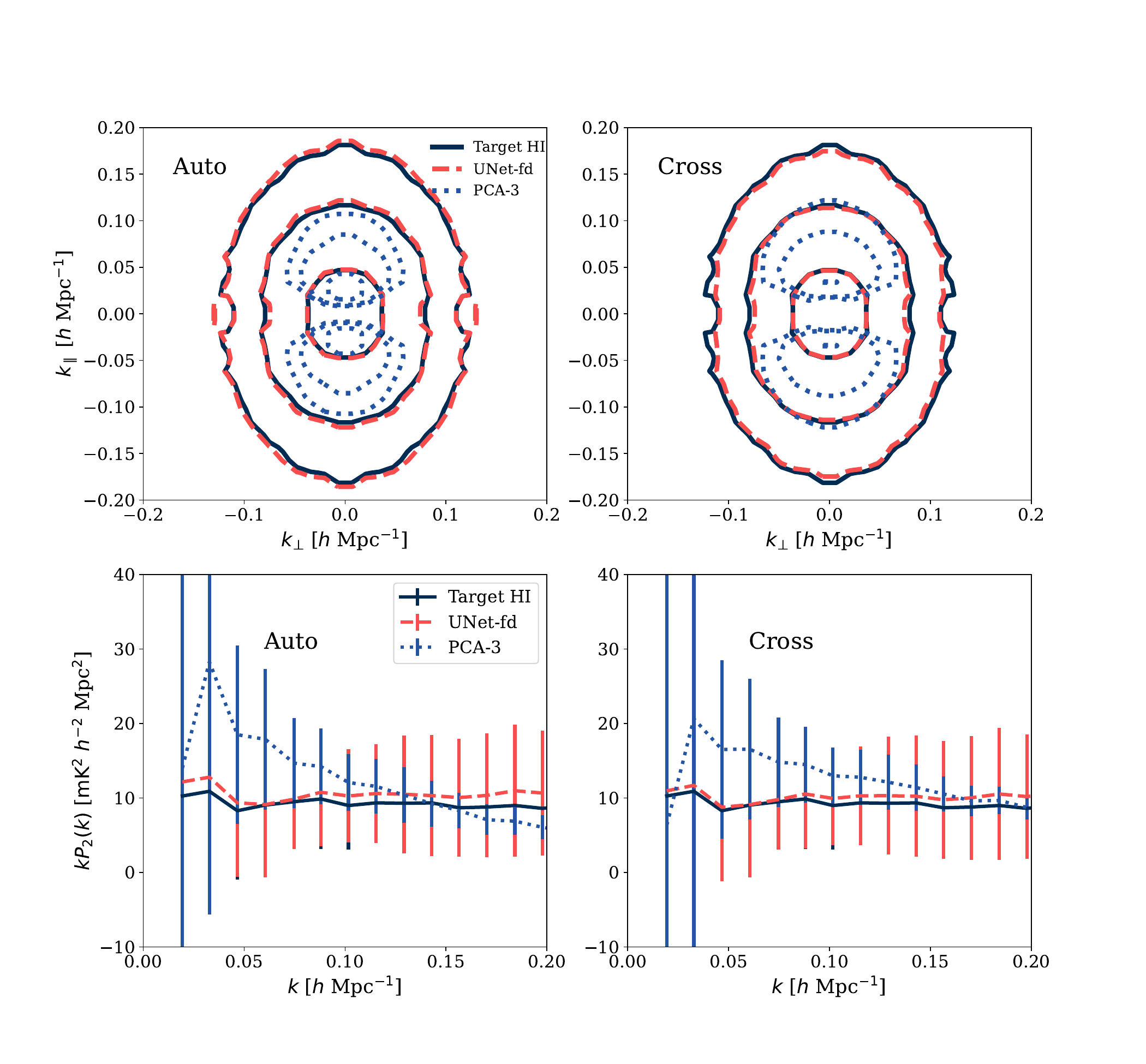}
    \caption{Same as in Fig.~\ref{fig:Pk2d_Hf}, but for Case II.}
    \label{fig:Pk2d_Hfb}
\end{figure*}

In order to gauge the accuracy of the reconstruction of the redshift-space distortions (RSD), we now compare $P(k_\bot, k_\parallel)$, as shown in the upper panels of  Fig.~\ref{fig:Pk2d_Hf}. The left and right panels correspond to the auto-correlation and cross-correlation power spectrum, respectively.  Black solid, red dashed, and blue dotted contours correspond to the true HI, the UNet-fd reconstruction, and the PCA-2 reconstruction, respectively. Note that the results are averaged from the 40 test cubes. In the case of UNet-fd, both auto-correlation and cross-correlation $P(k_\bot, k_\parallel)$ are clearly anisotropic with the ``elongated" feature in the central region, revealing the impact of the Kaiser effect on large scales (small k values). A comparison with the true $P(k_\bot, k_\parallel)$ shows that the UNet-fd reconstructed RSDs are overall very successful. In contrast, the PCA reconstruction shows significant signal reduction at small $|k_\parallel|$, which is always removed during the PCA foreground subtraction \citep{2015ApJ...815...51S}. However, such signal reduction is negligible in our method.

The Kaiser effect cause the quadrupole $P_2(k)$ to deviate significantly from zero, which is usually used to constrain the growth rate of the cosmic structure. The lower panels of Fig.~\ref{fig:Pk2d_Hf} compare $P_2(k)$ for both auto-correlation (left panel) and cross-correlation (right panel). Black solid, red dashed, and blue dotted lines correspond to the true HI, the UNet-fd reconstruction, and the PCA-2 reconstruction, respectively. Error bars indicate the $\pm1$ variance among the 40 test sets. As shown, the UNet-fd reconstructed $P_2(k)$ has deviations from zero, which is consistent with the target truth. Although there is a large uncertainty due to the cosmic variance in our small-volume samples, we can still observe a discrepancy between the PCA reconstructed and the true $P_2(k)$.

As a result, while just taking into account the foreground-contaminated 21-cm signal, our UNet-fd method can reliably remove the foreground and consistently reconstruct the redshift-space brightness temperature field of the cosmic HI.

\begin{figure*}
    \centering
    \includegraphics[trim=4cm 4cm 2cm 2cm, clip=True,width=2.1\columnwidth]{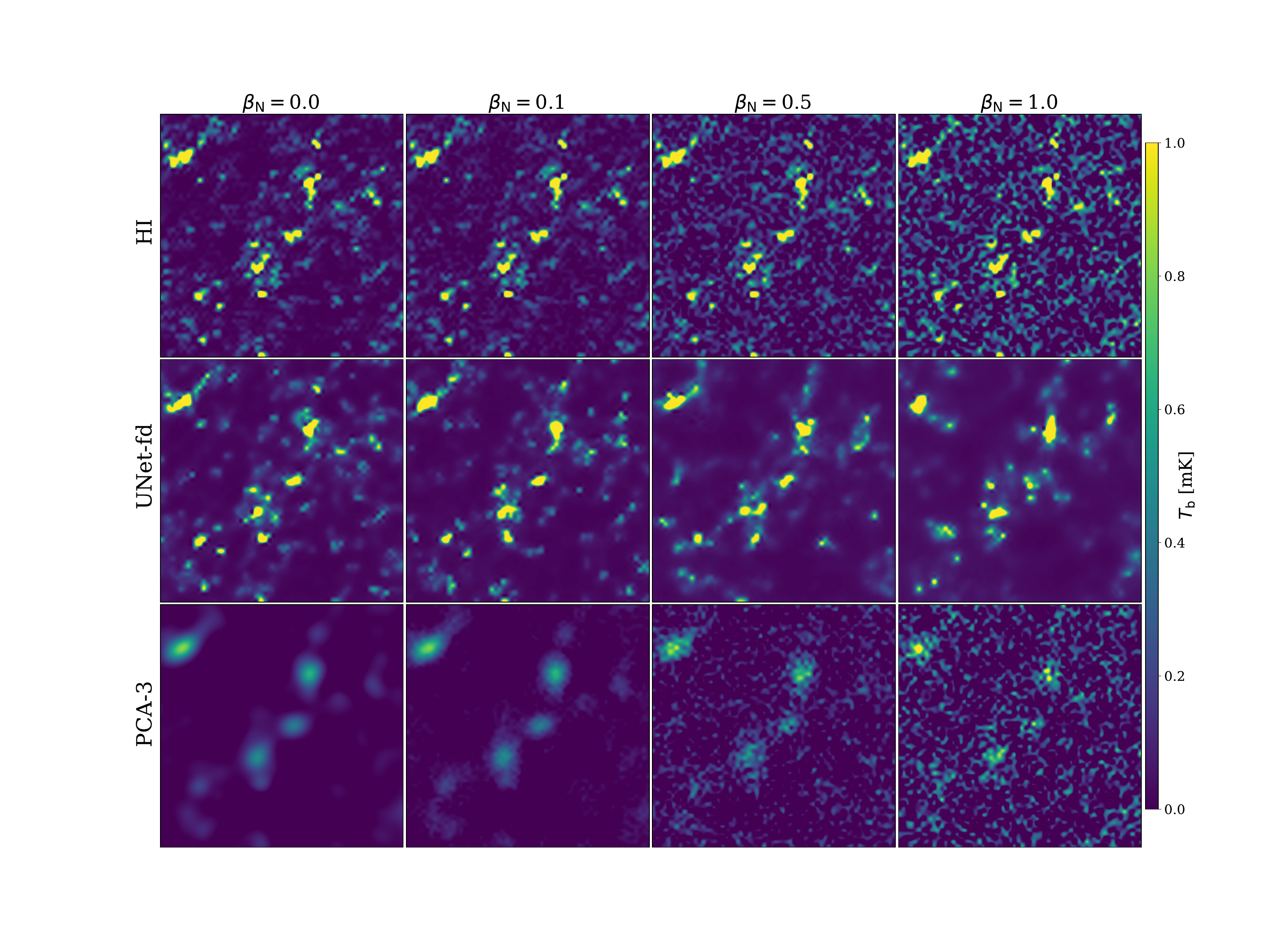}
    \caption{Comparison of the brightness temperature field for Case III in introducing thermal noise. Different columns correspond to different noise levels, as indicated at the top of each column. The first row presents the target HI with different levels of added noise (where the HI with $\beta_\rmns=0.0$ is the same target HI as in Fig.~\ref{fig:img_Hf} and \ref{fig:img_Hfb}). The second and third rows correspond to the UNet-fd and PCA-3 reconstructions respectively, as indicated to the left of each row. }
    \label{fig:img_Hfbn}
\end{figure*}
\begin{figure*}
    \centering
    \includegraphics[trim=0cm 0cm 0cm 0cm, clip=True,width=2\columnwidth]{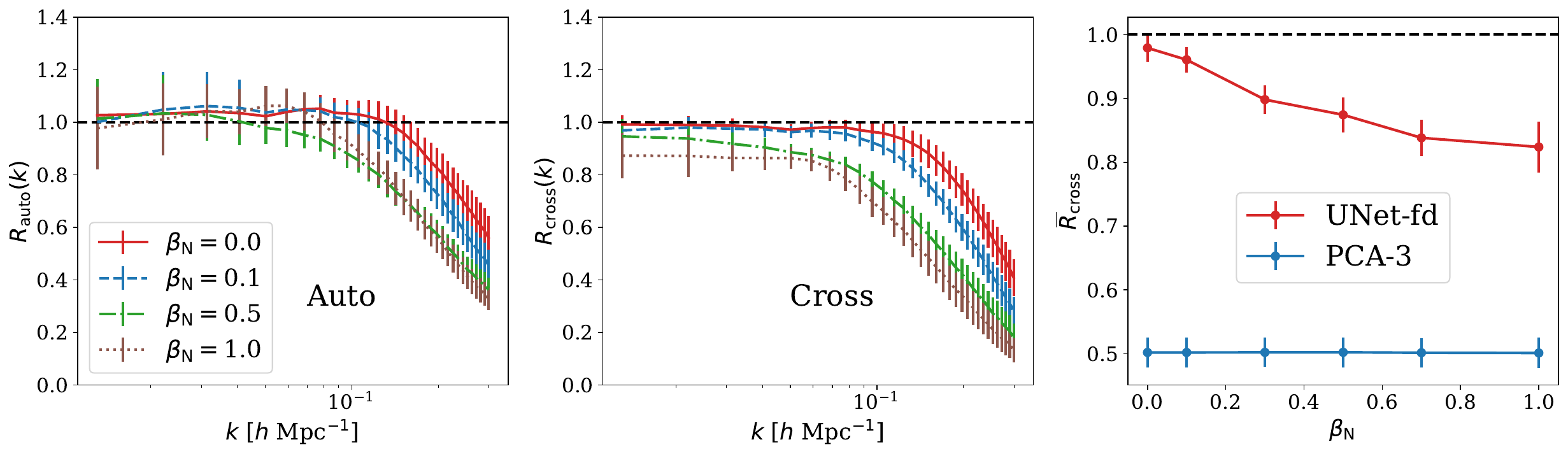}
    \caption{ Results of $R_\mathrm{auto}(k)$ (left) and $R_\mathrm{cross}(k)$ (middle) for Case III are shown using the UNet-fd method, with varying noise levels as indicated. The values for $\beta_\mathrm{ns}=0$ are the same as the UNet-fd results in Fig.~\ref{fig:Pk1d_Hfb}. The right panel shows a comparison of the mean cross-correlation ratio, $\bar{R}_\mathrm{cross}(\beta_\rmns)$, between UNet-fd (red) and PCA-3 (blue).} The value of $\bar{R}_\mathrm{cross}$ is calculated by averaging $R_\mathrm{cross}(k)$ from the values over scales $0<k<0.1\hmpc$.
    \label{fig:Pk1d_Hfbn}
\end{figure*}

\subsection{Case II: Considering beam effect}
It is shown that the foreground contamination can be removed efficiently depending on the feature of frequency correlation. However, the systematic beam effect reduces the correlation feature and causes non-smooth foreground residual, which challenges the detection of the HI signal. In this subsection, we validate our method for Case II.

Fig.~\ref{fig:img_Hfb} shows the visual comparison of the brightness temperature field. From left to right, the first panel shows the target HI field, which is the same as the one in Fig.~\ref{fig:img_Hf}. To demonstrate the effectiveness of the deconvolution, the second panel presents the target HI map convolved with the same beam as the corresponding foreground-contaminated map according to Eq.~(\ref{eq:beam}). For UNet-fd, the recovered temperature fluctuations nicely match the correspondence in the target field. Additionally, a noticeable deconvolution effect can be observed when comparing it to the second panel, indicating that the U-Net has learned to remove the beam effect. However, it has limited performance in predicting the detailed shape, suggesting that the beam effect is not perfectly eliminated. Comparatively, the PCA performs substantially worse, which always results in the signal being subtracted, especially in the beam case.

Fig.~\ref{fig:HPR_Hfb} shows the temperature distribution (left panel), pixel-to-pixel temperature comparison (middle panel), and the angular cross-correlation ratio as a function frequency channel (right panel). All panels are the same as those in Fig.~\ref{fig:HPR_Hf} but for testing the additional beam effect. In the middle panel, the two contours of each case encompass 90\% and 95\% of the grid cells, respectively. The UNet-fd recovered temperature distribution still follows consistently with the target and shows an unbiased pixel-to-pixel relation. However, the beam effect causes a larger scattering in pixel-to-pixel relationships and a 30\% reduction in angular cross-correlation ratios. The PCA reconstruction, in comparison, has a drastically reduced temperature distribution, biased pixel-to-pixel relationships, and cross-correlation ratios underestimated by around 60\%.
% The target HI temperature distribution is nevertheless consistently followed by the UNet-fd reconstruction, which shows an unbiased pixel-to-pixel relation.

Fig.~\ref{fig:Pk1d_Hfb} shows $R_\mathrm{auto}(k)$ (left panel) and $R_\mathrm{cross}(k)$ (right panel). Red solid and blue dashed lines correspond to UNet-fd and PCA reconstruction, respectively. In UNet-fd reconstruction, both values of $R_\mathrm{auto}(k)$ and $R_\mathrm{cross}(k)$ are consistent with unity at the $1\sigma$ level over the scales $k < 0.1\hmpc$. The underestimation of $R_\mathrm{cross}(k)$ at scale $k>0.1\hmpc$ is caused by the beam effect in comparison to the beam-free result of Fig.~\ref{fig:Pk1d_Hf}. Though, the overall reduction of the cross-correlation power spectrum at $k=0.2\hmpc$ and $0.3\hmpc$ are at 10\% and 30\% levels, respectively. The PCA, in contrast, exhibits large fluctuations in $R_\mathrm{auto}(k)$ and a reduction in $R_\mathrm{auto}(k)$, indicating an uncleaned recovery with lingering residual foreground and signal loss.

To test the RSD reconstruction, Fig.~\ref{fig:Pk2d_Hfb} shows the comparisons of $P(k_\bot, k_\parallel)$ (upper panels) and $P_2(k)$ (lower panels). All panels are the same as those in Fig.~\ref{fig:Pk2d_Hf} but for considering additional beam effect. In this case, both UNet-fd's auto-correlation and cross-correlation $P(k_\bot, k_\parallel)$ are in good agreement with the target truth, except for a few discrepancies of auto-$P(k_\bot, k_\parallel)$ at large $|k_\parallel|$. The RSDs are also recovered successfully with the anisotropic Kaiser features apparent. As expected, the quadrupole $P_2(k)$ deviates from zero and agrees with the target truth. For both the auto- and cross-correlation, the PCA results show a highly atypical shape of $P(k_\bot, k_\parallel)$ and unjustifiable amplitude of $P_2(k)$.

Therefore, even when taking into account the systematic beam effect, UNet-fd is still effective in removing the foregrounds.

\subsection{Case III: Introducing thermal noise}

Along with foregrounds and beam effects, we also take into account the instrumental thermal white noise as the Case III. As described in Sec.\ref{sec:data_ns},  we are not referring to any specific experiment for HI but rather simulating various levels of uncorrelated Gaussian noise. In this subsection, we verify how the noise affects our cleaning procedure, and look for a guideline threshold level for possible future  configurations of intensity mapping experiments.

Fig.~\ref{fig:img_Hfbn} shows the visual comparison of the brightness temperature field. Different columns correspond to different noise levels: $\beta_\rmns=0.0$, $\beta_\rmns=0.1$, $\beta_\rmns=0.5$, and $\beta_\rmns=1.0$, as indicated at the top of each column. To demonstrate the effectiveness of noise elimination, the first row presents the target HI with different levels of added noise (where the HI with $\beta_\rmns=0.0$ is the same target HI as in Fig.~\ref{fig:img_Hf} \& \ref{fig:img_Hfb}). The second and third rows correspond to the UNet-fd and PCA-3 reconstructions respectively, as indicated to the left of each row. Upon comparing the first two rows, it becomes evident that UNet-fd successfully restores the high-temperature structures while eliminating random noise. This highlights its ability to effectively separate the signal from Gaussian noise. The expected outcome from UNet is based on the distinct spatial distribution patterns of HI and noise. The HI demonstrates scale-dependent clustering, while the noise is dispersed in an uncorrelated random manner. However, when it comes to recovering the low-temperature structure, especially under high levels of noise, U-Net's capability to separate the signal is somewhat limited. Additionally, the noise variance of the difference maps is twice that of the original frequency maps, further compromising the accuracy of the U-Net reconstruction of the 21-cm signal. Nonetheless, the reconstructions obtained through PCA yield even poorer results, as the method is unable to effectively subtract the noise.

Fig.~\ref{fig:Pk1d_Hfbn} shows $R_\mathrm{auto}(k)$ (left panel) and $R_\mathrm{cross}(k)$ (middle panel) for the UNet-fd reconstructions. Different color lines are shown for different-level noise as indicated, where lines of $\beta_\rmns=0$ are the same as the UNet-fd results in Fig.~\ref{fig:Pk1d_Hfb}. The results of $R_\mathrm{auto}(k)$ are all near to unity at $1\sigma$ level over large scales, but have reductions at small scales (e.g. $k > 0.1\hmpc$). For the cross-correlation case, it reveals underestimates throughout all scales.

In the right panel of Fig.~\ref{fig:Pk1d_Hfbn}, we show the mean cross-correlation ratio, $\bar{R}_\mathrm{cross}(\beta_\rmns)$, for a comparison of UNet-fd (red) versus the PCA-3 (blue). We compute $\bar{R}_\mathrm{cross}$ by averaging $R_\mathrm{cross}(k)$ over scales $0<k<0.1\hmpc$. The UNet-fd result shows an inverse relation between $\bar{R}_\mathrm{cross}$ and $\beta_\rmns$, providing a practical guideline. It shows that the mean cross-correlation ratio remains above 0.8 when the variance of the thermal noise is smaller than or equal to that of the HI signal. In this case, the UNet-fd method reaches a limitation, leading to a systematic loss of the 21-cm signal. On the other hand, the PCA results exhibit significantly lower cross-correlation ratios, with values being around 0.5. This highlights the superior performance of the UNet-fd method compared to PCA in terms of cross-correlation ratios.

\section{Summary and conclusion}\label{sec:summary}

In this study, we propose the frequency-difference technique for pre-processing the data in U-Net deep learning for 21-cm foreground removal. We find that this technique dramatically suppresses the dynamic range of the foreground amplitudes by $2-3$ orders of magnitude. This improvement significantly enhances the performance of U-Net in foreground removal. Additionally, it circumvents the uncertainty associated with the traditional PCA pre-processing method, which involves selecting the number of eigenmodes and avoids the 21-cm signal loss often associated with PCA. Our tests reveal that the frequency-difference technique and U-Net are primarily employed to eliminate smooth and non-smooth foreground components, respectively.

We first verify our method using the 21-cm temperature including HI and foregrounds only. By conducting a series of comparison tests, including image comparison, pixel-to-pixel temperature relation, temperature distribution, and power spectrum, we find that our method can reliably remove the foreground. Both auto-correlation and cross-correlation power spectrum ratios agree excellently with unity over the ranges of scales $k<0.3\hmpc$, indicating a successful recovery of the HI clustering and remarkable concordance between the reconstructed and the true HI temperature field. Further considering the systematic beam effects, our reconstruction has auto-correlation and cross-correlation power spectrum ratios consistently at the $1\sigma$ level over the scales $k < 0.1\hmpc$. Beam effects smear out the small-scale clustering with 10\% and 30\% reduction of the cross-correlation power spectrum at $k=0.2\hmpc$ and $0.3\hmpc$, respectively. Moreover, the RSD effects are also reconstructed overall very successfully with the anisotropic Kaiser features apparent.

As a comparison, our method outperforms PCA, which has a drastically reduced temperature distribution, biased pixel-to-pixel temperature relationships, and cross-correlation ratios underestimated by around 60\%. PCA also fails to reconstruct the RSD structure, which has $P(k_\bot, k_\parallel)$ with a highly atypical shape and $P_2(k)$ with unjustifiable amplitude.

We eventually investigate the impact of instrumental thermal noise on the cleaning procedure and find that, with the help of the U-Net network, our approach can separate the signal from the noise.  By simulating various levels of Gaussian uncorrelated noise, we present an inverse relationship between the mean cross-correlation ratio $\bar{R}_\mathrm{cross}$ and the noise level $\beta_\rmns$. It suggests that if the variance level of the thermal noise is smaller than or equal to the HI signal, our method should be capable of obtaining a signal recovery with a cross-correlation ratio greater than approximately 0.8, surpassing the PCA result of 0.5. This information could be used as a reference for applying our method in practical experiments.

The large dynamic range of the foreground amplitude limits the deep learning method, which is now being applied in the 21-cm foreground removal. PCA preprocessing could be a solution, but it cannot solve the issue of HI signal loss. As a result, the frequency difference can considerably enhance network performance by reducing the amplitude range of the smooth foreground components and helping in the prevention of HI loss.\\\\

\section*{Acknowledgments}
This work is supported by the National SKA Program of China (2022SKA0110200, 2022SKA0110202, 2020SKA0110402, 2020SKA0110401), the National Key R\&D Program of China (2020YFC2201600), the National Natural Science Foundation of China (12103037, 12203038,11890691), the Fundamental Research Funds for the Central Universities (XJS221312), the Natural Science Basic Research Program of Shaanxi (Program No. 2022JQ-049), the Guangdong Basic and Applied Basic Research Foundation (No. 2021A1515110057),
the China Manned Space Project with No. CMS-CSST-2021 (A02, A03, B01), the Major Key Project of PCL, and the 111 project No. B20019. This work is supported by High-Performance Computing Platform of Xidian University and the Gravity Supercomputer at the Department of Astronomy, Shanghai Jiao Tong University. We also wish to acknowledge the Beijing Super Cloud Center (BSCC) and Beijing Beilong Super Cloud Computing Co., Ltd (\url{http://www.blsc.cn/}) for providing HPC resources that have significantly contributed to the research results presented in this paper.

\nocite{*}

\appendix*
\section{EXPLANATION ON USING UNET TO EXTRACT THE SINGLE-BAND HI FROM THE DIFFERENCE MAP}

\begin{figure*}
    \centering
    \includegraphics[trim=0cm 0cm 0cm 0cm, clip=True,width=1.5\columnwidth]{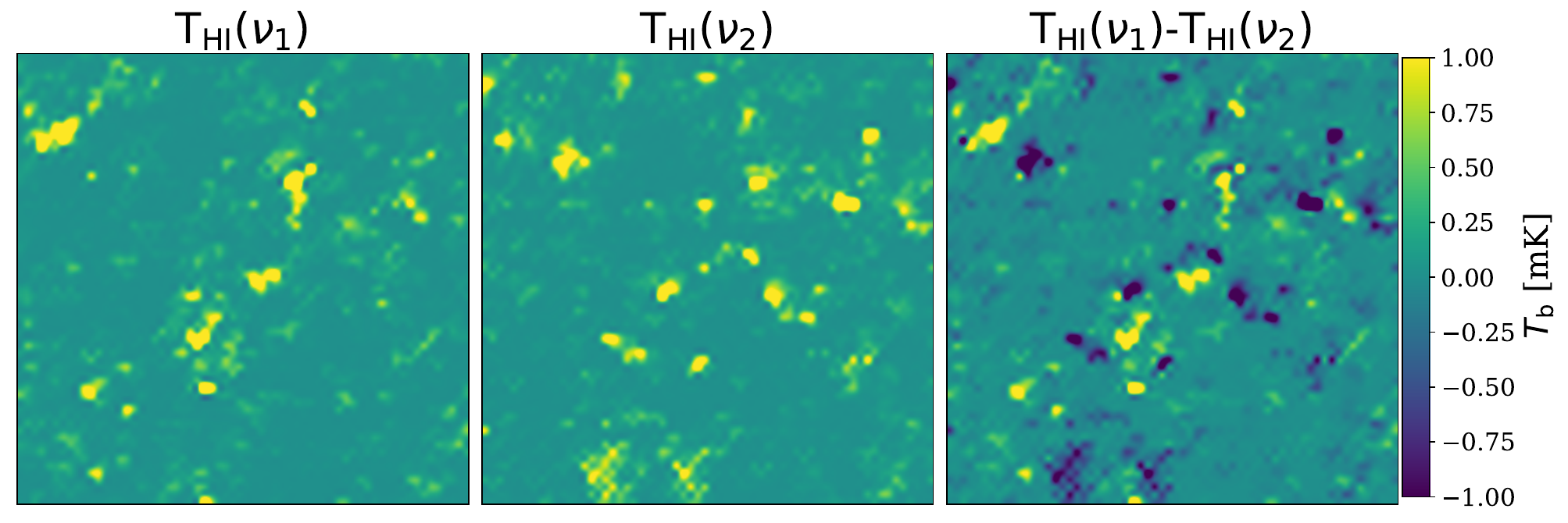}
    \caption{The pure HI temperature map at $\nu_1=1130$ MHz (left panel), $\nu_2=1132$ MHz (middle panel), and their corresponding difference map, $T_\text{HI}(\nu_1)-T_\text{HI}(\nu_2)$ (right panel). The $\nu_1$ slice is taken to be same as the target HI in Fig.~\ref{fig:img_Hf}, which is randomly selected from the dataset.}
    \label{fig:img_v1v2diff}
\end{figure*}

\begin{figure*}
    \centering
    \includegraphics[trim=0cm 0cm 0cm 0cm, clip=True,width=0.8\columnwidth]{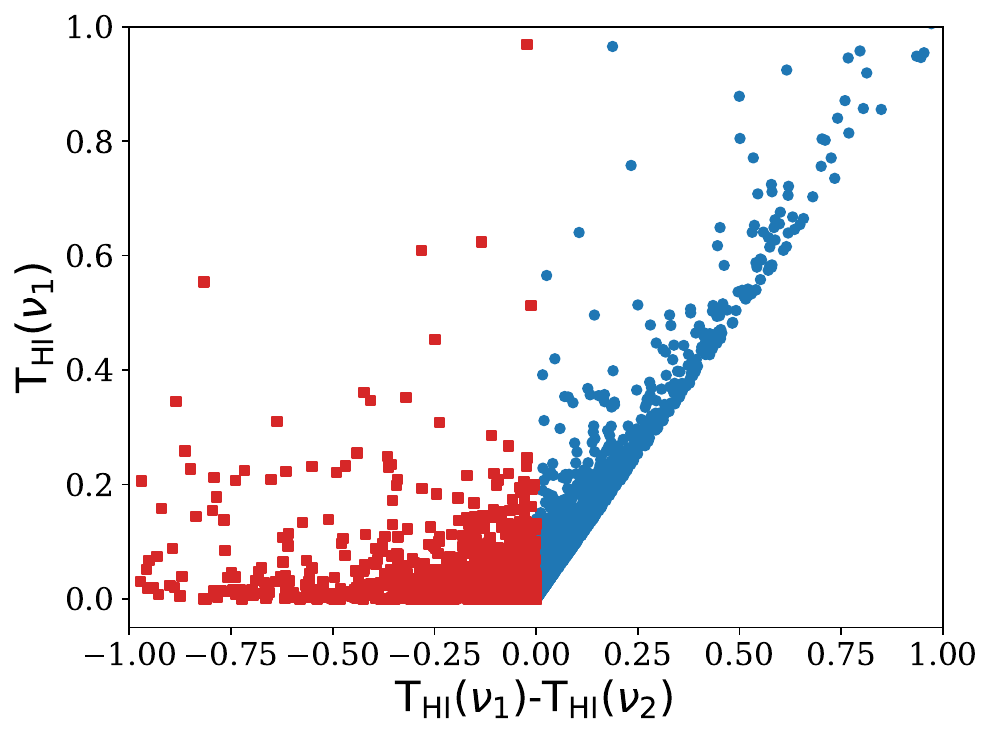}
    \caption{Pixel-to-pixel temperature relation, $T_\text{HI}(\nu_1)$ versus $T_\text{HI}(\nu_1)-T_\text{HI}(\nu_2)$. The pixels are selected from the slices shown in Fig.~\ref{fig:img_v1v2diff}. The blue circles and red squares indicate positive and negative values on the $x$-axis, respectively.}
    \label{fig:T_v1todiff}
\end{figure*}

As an illustration, Fig.~\ref{fig:img_v1v2diff} shows the pure HI temperature map at $\nu_1=1130$ MHz (left panel), $\nu_2=1132$ MHz (middle panel), and their corresponding difference map, $T_\text{HI}(\nu_1)-T_\text{HI}(\nu_2)$ (right panel). The difference map retains a significant number of structures from both frequency bands,  with most positive values corresponding to $\nu_1$ signals and most negative values corresponding to $\nu_2$ signals. It should be noted that the pure HI signals have a minimum value of zero in our dataset. Fig.~\ref{fig:T_v1todiff} shows the pixel-to-pixel temperature relation between $T_\text{HI}(\nu_1)$ and $T_\text{HI}(\nu_1)-T_\text{HI}(\nu_2)$. The blue circles reveals a significant positive correlation between $T_\text{HI}(\nu_1)$ and $T_\text{HI}(\nu_1)-T_\text{HI}(\nu_2)$, indicating the pixels associated with the $T_\text{HI}(\nu_1)$ map. In contrast, the red squares exhibit a much lower level of correlation, suggesting the pixels from the $T_\text{HI}(\nu_2)$ map. It is expected that such relationships will enable the UNet model to distinguish between $T_\text{HI}(\nu_1)$ and $T_\text{HI}(\nu_2)$, as well as extract the targeted single-band HI signals from the difference map. However, the presence of biased pixels within the blue circles and red squares is a result of mutual cancellation by the overlapping region in the difference map between $T_\text{HI}(\nu_1)$ and $T_\text{HI}(\nu_2)$. This may introduce some error when attempting to recover the targeted HI signals.

From another perspective, if the signals in the frequency-difference maps are purely random and lack any specific structure or features, such as white noise, it would be impossible to recover either of the signal maps from their difference maps using any method, including AI. This is because there are far more degrees of freedom that need to be recovered than the number of dimensions in the data. However, in our case, the HI signal does exhibit some structure, and the number of degrees of freedom for major features (eigenmodes) contained in the signal map is less than the dimension of the frequency-difference map. Therefore, this is the reason why we were able to effectively recover the HI maps for each frequency from the frequency-difference maps.
%
% %
% %

\bibliography{apssamp}% Produces the bibliography via BibTeX.

\end{document}